\documentclass{article} 
\usepackage{iclr2027_conference,times}

\usepackage{amsmath,amsfonts,bm}

\def\eqref#1{equation~\ref{#1}}

\def\1{\bm{1}}

\def\vb{{\bm{b}}}

\def\vp{{\bm{p}}}
\def\vq{{\bm{q}}}

\def\vx{{\bm{x}}}

\DeclareMathAlphabet{\mathsfit}{\encodingdefault}{\sfdefault}{m}{sl}
\SetMathAlphabet{\mathsfit}{bold}{\encodingdefault}{\sfdefault}{bx}{n}

\usepackage{hyperref}
\usepackage{url}
\usepackage{graphicx}
\usepackage{booktabs}
\usepackage{colortbl}
\usepackage{pifont}
\usepackage{wrapfig}
\usepackage{xcolor}
\usepackage{enumitem}

\definecolor{darkgreen}{RGB}{198,239,206}
\definecolor{pippurple}{RGB}{112,48,160}

\providecommand{\cmark}{\ding{51}}
\providecommand{\xmark}{\ding{55}}

\title{Fourier-Latent Diffusion for Constrained Generation of Triply Periodic Minimal Surfaces}

\author{Shu Yan, Ruihong Cen \& Bohan Wang \\
Department of Computer Science, National University of Singapore
}

\iclrfinalcopy 
\begin{document}

\maketitle
\lhead{Preprint}

\begin{figure*}[h]
    \centering
    \includegraphics[width=1\textwidth]{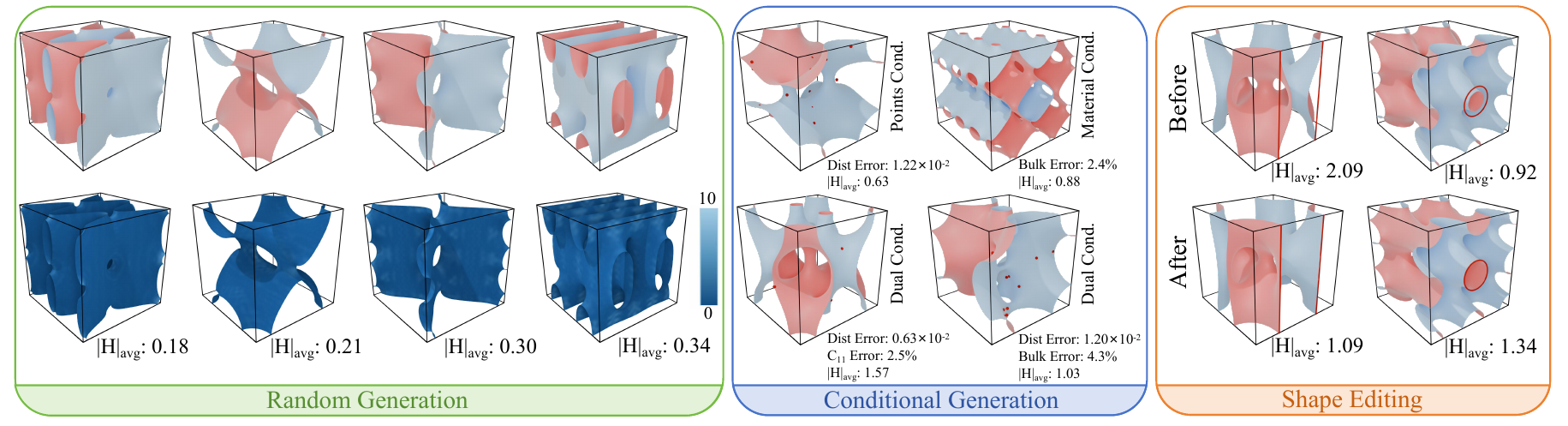}
    \caption{\textbf{Our framework enables diverse and near-minimal TPMS generation given user requirements.}
    \textbf{Left:} unconditional TPMS samples.
    \textbf{Middle:} generation conditioned on sparse points, selected elastic properties, or both.
    \textbf{Right:} local editing.
    Across all these tasks, the generated TPMS structures maintain low mean curvature over the entire surface.}
    \label{fig:teaser}
\end{figure*}

\begin{abstract}
We present a generative framework for the constrained design of $D_{2h}$-symmetric triply periodic minimal surfaces (TPMS) with low residual mean curvature.
Existing TPMS design pipelines either explore a limited set of canonical analytical families or rely on computationally expensive numerical procedures, 
making it difficult to efficiently generate diverse candidates under geometric and/or mechanical requirements.
Our approach learns a distribution of solver-generated TPMS in a compact Fourier space 
where periodicity and $D_{2h}$ symmetry are guaranteed by construction.
This representation eliminates the need for a learned geometric decoder and enables the training of an effective diffusion model that generates low-mean-curvature candidates 
conditioned on sparse geometric constraints, selected homogenized elastic properties, or their combination.
A subsequent coefficient-space refinement further reduces the residual mean curvature.
Experiments show that our approach outperforms trigonometric, SDF-based, and standard Fourier-space baselines 
and enables controllable, high-quality generation under geometric and low-dimensional mechanical constraints.
Overall, the proposed framework provides a compact and efficient design space for generating near-minimal periodic structures under user-specified requirements.

\end{abstract}

\section{Introduction}
\label{sec:introduction}

Triply periodic minimal surfaces (TPMS) are periodic surfaces 
with zero mean curvature that repeat along three independent spatial directions.
The near-zero mean curvature of TPMS is not only a defining geometric property, 
but also has important implications for their physical performance.
In particular, TPMS can attain the optimal bulk modulus when the thickness approaches zero, 
establishing a direct connection between zero mean curvature and mechanical efficiency~\citep{torquato2003optimal,zhang2025asymptotic}.
Together with their smooth and bicontinuous geometry, 
high surface-area-to-volume ratio, 
and interconnected transport pathways,
these properties have motivated broad applications of TPMS 
in architected materials, heat and mass transfer, and related engineering systems~\citep{feng2022triply}.
For such applications, the design problem extends beyond reproducing a small set of familiar TPMS families: 
designers often seek geometrically diverse candidates that satisfy prescribed spatial and functional requirements.
For example, a structure may be required to pass through selected locations while simultaneously achieving a target bulk modulus.
Addressing such problems requires a design space that can explore diverse TPMS morphologies while preserving low mean curvature and supporting quantitative geometric and mechanical constraints.

Existing approaches provide complementary but incomplete solutions.
A widely used strategy represents canonical TPMS families, such as the Schwarz Primitive, Diamond, and Gyroid surfaces, using a small number of trigonometric terms~\citep{al2021mslattice,jones2021tpms,hsieh2020minisurf}.
These representations are simple, efficient, and convenient for modeling and fabrication, but their restricted parameterizations cover only a limited portion of the broader TPMS design space and generally introduce non-negligible residual mean curvature.
Because mean curvature is directly related to both minimal-surface fidelity and the mechanical optimality, such approximation errors are not merely geometric artifacts and may compromise the desirable physical properties of TPMS.
Procedural TPMS construction provides more accurate surfaces and access to a substantially broader family by numerically solving minimal-surface problems associated with different boundary configurations~\citep{makatura2023procedural}.
However, repeatedly specifying boundary configurations and solving the corresponding numerical problems is computationally expensive and does not directly provide a mechanism for efficiently sampling new candidates under geometric and mechanical requirements.

We address this gap within a $D_{2h}$-symmetric design space by learning the distribution of high-quality, solver-generated TPMS.
Our key idea is to represent these surfaces in a symmetry-reduced Fourier space in which periodicity and coordinate-reflection symmetry are enforced exactly by construction, rather than learned approximately from data.
By explicitly incorporating $D_{2h}$ symmetry into the basis, the representation removes symmetry-related Fourier redundancy and provides a compact coefficient tensor for generative modeling.
At the same time, these coefficients define a continuous implicit field whose spatial derivatives can be evaluated analytically, enabling direct surface extraction and curvature-based refinement without requiring an additional learned geometric decoder.
Then, we train a conditional diffusion model in this compact space to capture the geometric distribution of solver-generated TPMS.
The model supports unconditional sampling as well as conditional generation from sparse geometric constraints and selected homogenized elastic properties.
Periodicity and $D_{2h}$ symmetry are guaranteed by the representation, 
whereas near-minimal curvature is learned from the solver-generated data distribution.
A subsequent coefficient-space optimization further reduces residual mean curvature 
while remaining within the same periodic and symmetry-constrained representation.
Although the resulting surfaces are not exact zero-mean-curvature minimal surfaces, 
they provide near-minimal TPMS candidates with lower residual mean curvature than 
the evaluated trigonometric representation and the SDF- and standard-Fourier-space generative baselines.

\begin{figure}[t]
\centering
\includegraphics[width=0.9\linewidth]{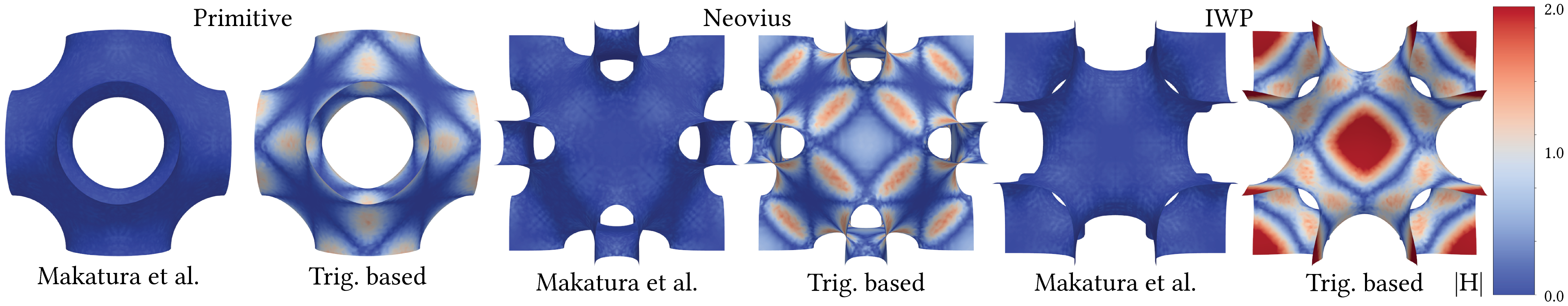}
\vspace{-0.3cm}
\caption{\textbf{TPMS represented by low-order trigonometric fields exhibit large absolute mean curvature.}
We compare three classical TPMS: Primitive, Neovius, and IWP constructed using the accurate method from~\citet{makatura2023procedural}
and low-order trigonometric fields, respectively. 
The latter exhibits substantially higher absolute mean curvature.} 
\label{fig:tpms-vs-loop}
\end{figure}

Our contributions are threefold.
\textbf{First}, we construct a dataset of 18,704 aligned TPMS candidates by combining procedural minimal-surface construction with boundary-loop enumeration, filtering, and canonicalization.
\textbf{Second}, we develop a conditional generative framework in a compact $D_{2h}$-invariant Fourier space that guarantees periodicity and symmetry by construction, captures low-curvature TPMS morphologies through learning, and supports coefficient-space curvature refinement.
\textbf{Third}, we evaluate unconditional generation and conditional generation under sparse point constraints, 
low-dimensional elastic-property targets, and their combination, 
together with comparisons that examine the effects of the representation and curvature refinement.

Collectively, the proposed framework broadens the accessible design space of $D_{2h}$-symmetric TPMS while maintaining low residual mean curvature and enabling inverse design under both geometric and mechanical requirements.
To the best of our knowledge, it is the first learned generative framework for TPMS design that jointly supports geometric and material-property conditioning while maintaining low residual mean curvature.

\section{Related Work}
\label{sec:related_work}
\noindent\textbf{Periodic and minimal-surface construction.}
Classical minimal-surface constructions and geometric solvers provide a foundation for TPMS modeling.
The Enneper--Weierstrass representation describes exact minimal surfaces when the associated period conditions are satisfied, 
but obtaining suitable data and enforcing global compatibility make broad enumeration difficult~\citep{meeks2011classical}.
Discrete methods recover particular families through specialized constructions~\citep{reitebuch2019discrete}, 
while Plateau-type solvers compute minimal surfaces from prescribed boundary~\citep{pinkall1993computing,wang2021computing}.
Our dataset builds on the procedural boundary-loop construction of \citet{makatura2023procedural}, 
which computes a minimal patch on a fundamental domain and extends it by symmetry.
Low-order trigonometric fields are the most used TPMS design tools~\citep{al2021mslattice,jones2021tpms,maskery2022flatt,hsieh2020minisurf}.
Weighted combinations further support hybridization and property variation~\citep{feng2021isotropic,khaleghi2021directional}, although these fields generally define approximations rather than exact minimal surfaces.
Figure~\ref{fig:tpms-vs-loop} illustrates residual-curvature differences
between solver-derived examples and low-order trigonometric approximations.
Other approaches construct shell or surface architectures from labyrinths and boundary parameterizations~\citep{akbari2020geometry,akbari2022strut,xu2023new}.
Our method also uses a finite Fourier representation, 
but learns its coefficient distribution from solver-generated data 
instead of restricting design to a small set of canonical formulas.

\noindent\textbf{Generative modeling.}
Three-dimensional generative models operate on representations ranging from point clouds and meshes to volumetric and continuous implicit fields~\citep{yang2019pointflow,liu2023meshdiffusion,zheng2022sdf}.
For high-resolution geometry, representation size affects both learning cost and reconstruction accuracy.
Latent diffusion addresses this issue through learned compression~\citep{cheng2023sdfusion,chou2023diffusion}, while transform-domain methods exploit wavelet, spectral, or cosine structure~\citep{hu2024neural,fan2025mesh,ning2024dctdiff}.
We follow the latter principle with a basis tailored to periodic, reflection-symmetric surfaces.
Its coefficient relations are imposed explicitly, so the denoiser need not learn redundant encodings of the same symmetry orbit.
The backbone uses established diffusion and transformer components~\citep{ho2020denoising,bao2023all}.
For generative inverse design, 
MIND~\citep{xue2025mind} generates tileable microstructures using a hybrid neural representation, and guided diffusion for voxel-based metamaterials~\citep{yang2026guided} targets specified elastic responses.
These methods address closely related inverse-design problems.
Our emphasis is the combination of sparse geometric conditions, mechanical targets, and thin shells with low residual curvature.
This specialization motivates the representation but does not make prior inverse-design methods inapplicable.

\begin{figure}
    \centering
    \includegraphics[width=0.85\linewidth]{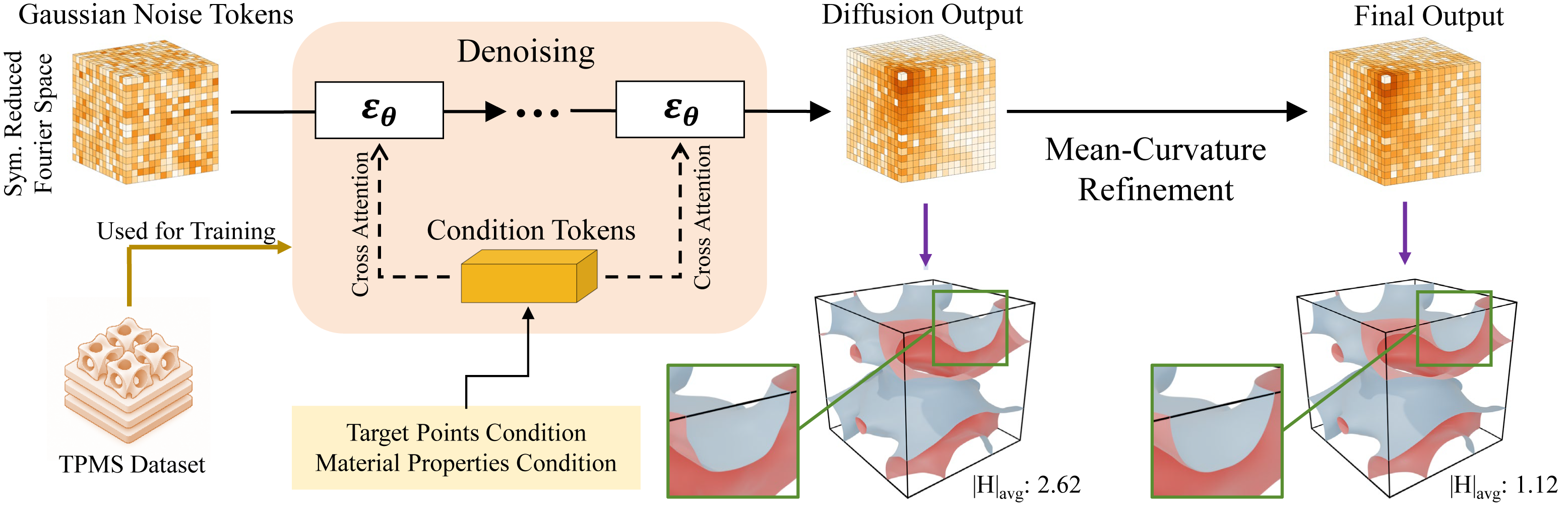}
    \vspace{-0.3cm}
    \caption{\textbf{Overview of our generation pipeline.}
    Starting from Gaussian noise in our $D_{2h}$-reduced space, 
    the shared denoising network $\varepsilon_\theta$ iteratively removes noise 
    while optionally incorporating point and/or material-property conditions as tokens through cross-attention.
    The denoised coefficients correspond to a surface with low mean curvature, 
    and our postprocessing step further reduces the residual curvature, 
    bringing the surface closer to satisfying the minimality requirement of TPMS.}    
    \label{fig:pipeline-full}
\end{figure}

\noindent\textbf{Learning Solution Fields of PDEs.}
Because minimal surfaces satisfy a zero-mean-curvature partial differential equation (PDE), 
our problem is related to learned solution fields, 
spanning PINNs~\citep{raissi2019physics}, 
geometry-informed methods~\citep{berzins2024geometry}, operator-learning approaches~\citep{li2020fourier,li2023geometry,wu2024transolver},
and physics-constrained generative modeling~\citep{chung2022diffusion,baldan2026physics}.
However, standard operator-learning methods typically assume known inputs 
such as boundary conditions, forcing terms, or geometric parameters, 
whereas TPMS boundary data are not specified a priori.

\section{Method}
\label{sec:method}
Our pipeline combines solver-based data construction, symmetry-reduced Fourier encoding, conditional diffusion, and curvature refinement (Figure~\ref{fig:pipeline-full}).
Let $\Omega=[0,L]^3$ denote a cubic periodic cell with side length $L$.
We use $L=1$.
The model operates on field coefficients rather than extracted meshes; the geometry is recovered as the zero set of the decoded field.

\subsection{Solver-Derived Dataset}
\label{sec:dataset}

\begin{figure}
    \centering
    \includegraphics[width=0.9\linewidth]{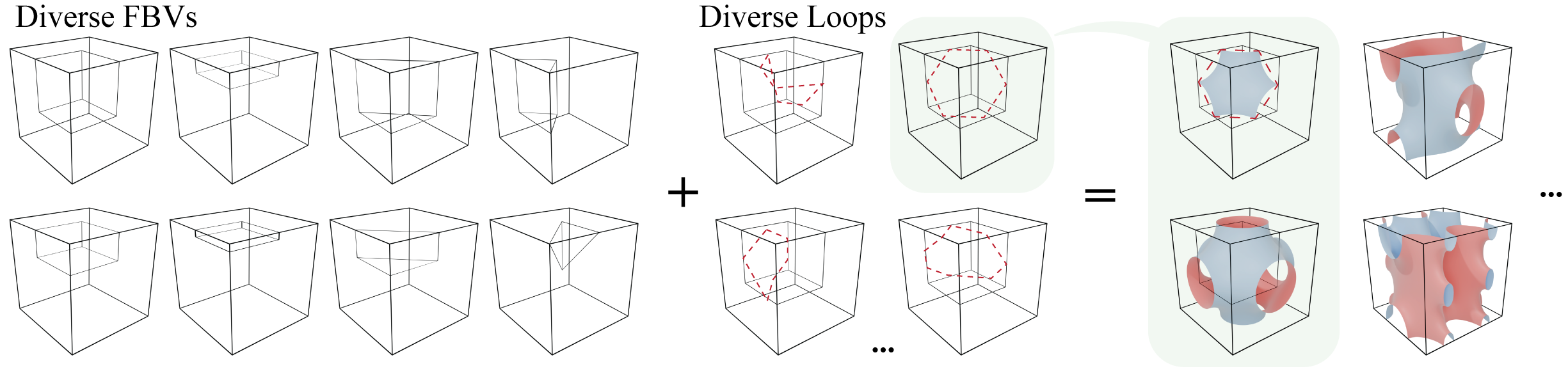}
    \vspace{-0.3cm}
    \caption{\textbf{TPMS dataset synthesis.}
    By exhaustively combining diverse admissible boundary loops with multiple mirrorable FBV types, 
    we generate a large set of geometrically diverse TPMS.}
    \label{fig:fbv}
\end{figure}

We use the procedural construction of \citet{makatura2023procedural} to 
compute free-boundary minimal-surface patches from 
admissible loops defined on mirrorable fundamental bounding volumes (FBVs).
An FBV is a subdomain whose reflected copies tile the periodic unit cell.
A boundary loop specifies a minimal-surface problem within the FBV, 
and reflecting the resulting patch across the FBV faces produces a TPMS with $D_{2h}$ symmetry.
We enumerate loops containing at most ten vertices across eight FBV types (Figure~\ref{fig:fbv}), 
retain successfully solved instances, and augment the resulting meshes using distinct canonical rotations of the unit cell.
This procedure explores diverse boundary configurations within the selected construction, 
rather than varying only a small number of parameters within a fixed surface family.
Each solved TPMS is converted into a signed distance field (SDF), 
using a consistent sign convention that requires the eight corners of the unit cell to have positive values.
We then symmetrize the fields, remove the constant Fourier component, 
and use the dominant Fourier coefficients to align representations that differ only by half-period shifts or global sign flips.
Duplicate geometries and candidates with an average absolute mean curvature above $7.5$ are discarded.
The resulting 18,704 aligned candidates are split into training and validation sets in a $9{:}1$ ratio after preprocessing.
Appendix~\ref{sec:dataset_details} provides further details on the encoding and canonicalization procedures.
Canonicalization is important because the same periodic geometry can admit multiple equivalent field representations: 
translating the unit-cell origin by half a period changes the signs of selected Fourier coefficients, 
whereas reversing the SDF sign leaves the zero level set unchanged.
Selecting a consistent representative therefore reduces this redundancy
and aligns the structures before learning.

After canonicalization and projecting the shape into our representation, 
the average absolute mean curvature decreases from $2.2$ for the raw meshes to $1.4$, 
while the geometric reconstruction error remains small.
Across the dataset, the mean absolute SDF difference is approximately $5\times10^{-4}$, 
and the 99th-percentile absolute SDF difference remains below $6.5\times10^{-3}$ for every structure (Figure~\ref{fig:stats-dataset}).
These measurements show that the processed dataset preserves the original geometry 
while exhibiting low residual mean curvature.
This provides the diffusion model with a consistent distribution of low-curvature, 
reflection-symmetric TPMS geometries from which to learn.

\subsection{Symmetry-Reduced Fourier Representation}
\label{sec:representation}
Directly learning a generative model over raw 3D shapes is challenging
because high-resolution field- or point-based representations are high-dimensional and computationally expensive.
Implicit neural representations such as DeepSDF~\citep{park2019deepsdf} can represent high-quality continuous geometry,
but accurate reconstruction typically requires dense spatial queries during training and surface extraction.
Likewise, image-based architectures such as convolutional U-Nets~\citep{ronneberger2015u} do not naturally enforce three-dimensional periodicity.
In contrast, periodic scalar fields admit Fourier expansions,
making their coefficients a natural representation for our data.
The additional reflection symmetry enables a substantial reduction in dimensionality.
After alignment, the fields are invariant under $p_x\mapsto L-p_x$, $p_y\mapsto L-p_y$, and $p_z\mapsto L-p_z$.
These three commuting reflections generate the eight-element group $D_{2h}$.
Invariance ties Fourier coefficients under independent sign changes of the coordinate frequencies, eliminating sine factors from the corresponding tensor-product basis.
We therefore write
\begin{equation}
    \begin{aligned}
    f(\mathbf p;\bm a)&=
    \sum_{h,k,\ell=0}^{K_{\max}}
    a_{hk\ell}
    \cos\!\left(\frac{2\pi h p_x}{L}\right)
    \cos\!\left(\frac{2\pi k p_y}{L}\right)
    \cos\!\left(\frac{2\pi \ell p_z}{L}\right),\\
    \mathcal S(\bm a)&=\{\mathbf p\in\Omega:f(\mathbf p;\bm a)=0\},
    \end{aligned}
    \label{eq:d2h-basis}
\end{equation}
where $K_{\max}$ is the largest retained frequency index in each coordinate and $\bm a$ is the real coefficient tensor.
Every coefficient tensor defines a periodic, $D_{2h}$-invariant field, regardless of whether it is obtained from the dataset, diffusion sampling, or refinement.

\begin{figure}
    \centering
    \includegraphics[width=0.9\linewidth]{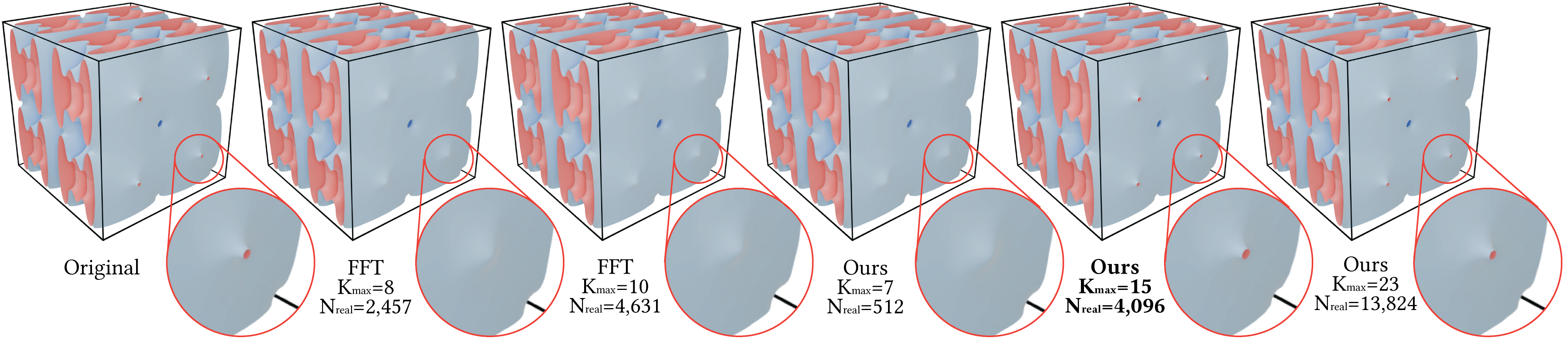}
    \vspace{-0.3cm}
    \caption{\textbf{Our latent representation is more compact than a standard Fourier representation.}
    Our \(D_{2h}\)-symmetric Fourier basis accurately captures fine TPMS features, including small holes, with 4,096 coefficients.
    In contrast, a standard Fourier representation with a similar coefficient budget fails to recover these features.
    Further increasing the number of coefficients in our basis yields negligible improvement.}
    \label{fig:different-basis}
\end{figure}

A general real Fourier expansion containing harmonics up to absolute frequency
$K_{\mathrm{max}}$ requires $(2K_{\mathrm{max}}+1)^3$ real degrees of freedom,
whereas our invariant basis uses only $(K_{\max}+1)^3$ coefficients.
For $K_{\max}=15$, the counts are $29{,}791$ and $4{,}096$, respectively, before fixing the constant coefficient.
In this work, we store the coefficients in a $16^3$ tensor and set $a_{000}=0$ during preprocessing.
The coefficients are obtained by combining symmetry-related standard Fourier coefficients.
Each nonzero cosine factor corresponds to a pair of positive- and negative-frequency complex Fourier modes.
Thus, if \(q\) denotes the number of nonzero indices among \(h,k,\ell\),
the corresponding cosine-product coefficient is obtained from the normalized DFT coefficient as
$a_{hk\ell}=2^q\operatorname{Re}(F_{hk\ell})$,
excluding the usual Nyquist-frequency special cases.
The detailed derivation is in Appendix~\ref{sec:basis_details}.
Figure~\ref{fig:different-basis} illustrates how the selected cutoff captures openings in a representative reconstruction.

This compactness results from restricting the function space rather than eliminating Fourier truncation error.
The decoded field is an implicit level-set function and need not remain a metric SDF.
Its zero set is smooth wherever $\nabla f\neq\mathbf 0$, 
but neither regularity nor zero mean curvature is guaranteed for arbitrary coefficients.
The $16^3$ tensor represents spectral coefficients, 
so the value and the curvature of the field can therefore be computed analytically.

\subsection{Conditional Diffusion}
\label{sec:diffusion}

\paragraph{Normalization and tokenization.}
Coefficient magnitudes vary substantially across frequency modes.
We apply the componentwise signed-logarithmic transform
$\tilde a_i=\operatorname{sign}(a_i)\log(1+|a_i|/\tau)$, where $\tau>0$ controls the scale,
followed by affine rescaling to obtain normalized coefficients $\hat{\bm{a}}\in[-1,1]^{(K_{\max}+1)^3}$.
The normalized $16^3$ tensor is partitioned into nonoverlapping $2\times2\times2$ blocks,
yielding $512$ tokens with eight coefficients each.
Let $\bm y_0=\mathcal T(\bm a)$ denote this complete encoding operation.
Its inverse undoes patchification and affine rescaling before applying the inverse signed-logarithmic transform, ensuring that geometry is always reconstructed from coefficients in the original field scale.

\paragraph{Denoiser and conditions.}
Inspired by~\cite{bao2023all}, our UViT-style denoiser linearly embeds the coefficient tokens into $256$ dimensions and processes them using four encoder blocks, one bottleneck block, and four decoder blocks with skip fusion and four attention heads.
A final linear head predicts noise in the original token space.
The diffusion timestep is provided as a time token.
For conditional generation, sparse point coordinates, selected homogenized elastic properties, or both are embedded as condition tokens and incorporated through cross-attention.
The condition is supplied at every reverse step, while cross-attention is omitted for unconditional generation.
This formulation allows the denoiser to incorporate design requirements while generating the full coefficient tensor, rather than modifying only a prescribed subset of geometric parameters.

\paragraph{Training.}
We use a variance-preserving forward process~\citep{song2020score} with the standard noise-prediction objective,
\begin{equation}
\begin{aligned}
    \bm y_t&=\alpha(t)\bm y_0+\sigma(t)\boldsymbol\epsilon,
    \qquad \boldsymbol\epsilon\sim\mathcal N(\mathbf 0,\mathbf I),\quad t\sim\mathcal U(0,1),\\
    \mathcal L_{\mathrm{diff}}&=
    \mathbb E_{\bm y_0,\boldsymbol\epsilon,t,\mathbf c}
    \bigl[\|\boldsymbol\epsilon-\boldsymbol\epsilon_\theta(\bm y_t,t,\mathbf c)\|_2^2\bigr].
\end{aligned}
\label{eq:diffusion_training}
\end{equation}
Here, $\alpha(t)=\exp\!\left(-\frac{1}{2}\int_0^t \beta(s)\,ds\right)$ and $\sigma(t)=\sqrt{1-\alpha(t)^2}$ determine the signal and noise amplitudes, $\beta(\cdot)$ is the noise schedule, and $\mathbf c$ denotes the optional condition.
The condition may encode sparse geometric constraints, target homogenized material properties, or both.
At each training step, a clean example is perturbed at a randomly sampled time, and the denoiser is trained to predict the injected Gaussian noise.
The final objective includes neither an Eikonal loss nor an explicit curvature penalty, as our ablation study shows that these terms are either unnecessary or detrimental to training (Appendix~\ref{sec:ablation-additional_training_losses}).
Instead, the low-curvature character of the geometry is learned directly from the data distribution.

\paragraph{Sampling, inversion, and editing.}
Generation starts from Gaussian tokens and uses deterministic DDIM updates~\citep{song2020denoising} to recover a clean coefficient representation.
Each update combines the current clean-sample estimate with the network-predicted noise; no additional randomness is introduced once the initial state is fixed.
Decoding with $\mathcal T^{-1}$ yields the coefficients used in Eq.~(\ref{eq:d2h-basis}).
Starting from an existing TPMS,
DDIM inversion maps its Fourier coefficients along a deterministic trajectory from the clean latent state to a noisier latent state.
Reversing the same trajectory approximately reconstructs the input shape,
allowing the noisy latent to serve as a starting point for controlled editing.
At each denoising step, we evaluate a differentiable constraint loss on the predicted clean surface
and take a gradient step with respect to the latent
to move the surface toward prescribed points or boundary curves before continuing the DDIM update.
For example, to force the generated surface to pass through a set of user-specified points $\mathcal{P}_{\mathrm{con}}$,
we use $\mathcal{L}_{\mathrm{geom}}=\frac{1}{|\mathcal{P}_{\mathrm{con}}|}\sum_{\mathbf{p}\in\mathcal{P}_{\mathrm{con}}}\left|f(\mathbf{p};\hat{\bm{a}}_0)\right|^2,$
where $\hat{\bm{a}}_0$ is obtained by inverse-normalizing $\hat{\bm{y}}_0$, the predicted clean latent at the current timestep.

\subsection{Post Curvature Refinement}
\label{sec:refinement}
Although our diffusion model already produces structures with sufficiently low mean curvature,
we further reduce residual curvature by directly optimizing the decoded coefficients.
At iteration $k$, we sample a point set $\mathcal P^{(k)}$ from the current surface
and perform a gradient-based update using
\begin{equation}
    \mathcal L_{\mathrm{ref}}^{(k)}(\bm a)=
    \mathbb E_{\mathbf p\sim\mathcal P^{(k)}}[H_\delta(\mathbf p;\bm a)^2]
    +\lambda_{\bm a}\mathbb E_{\mathbf p\sim\mathcal P^{(k)}}|f(\mathbf p;\bm a)|,
    \label{eq:curvature_refinement}
\end{equation}
where the regularized mean curvature is evaluated as
\begin{equation}
    H_\delta(\mathbf p;\bm a)=\frac12\nabla\cdot
    \left(\frac{\nabla f(\mathbf p;\bm a)}
    {\sqrt{\max\left(\|\nabla f(\mathbf p;\bm a)\|_2^2,\delta\right)}}\right),
    \qquad \delta>0.
    \label{eq:implicit_curvature}
\end{equation}
$\delta$ is a very small value to prevent division by zero.
The first term penalizes residual curvature, while the second discourages deviation from the reference surface.
The point set is resampled at each iteration rather than fixed on the initial level set,
so each update is evaluated relative to the preceding geometry.

The Fourier basis provides analytic first- and second-order spatial derivatives, avoiding finite-difference approximations.
The normalized-gradient curvature formula applies to any regular implicit field and does not require a metric SDF.
We use $8{,}192$ sampled vertices, $80$ optimization steps, and $\lambda_{\bm a}=0.1$.
Because refinement remains within the same coefficient space,
periodicity and reflection symmetry remain exact throughout optimization,
while the compact representation enables efficient refinement.

\section{Results}
\label{sec:results}

Our $7.8$M-parameter model is implemented in PyTorch~2.2.1 and trained for one million steps using AdamW~\citep{loshchilov2017decoupled,kingma2014adam}, a batch size of $128$, and a learning rate of $10^{-4}$.
Training takes approximately $30$ hours on two NVIDIA H200 GPUs.
On one H200, sampling takes $2.1$\,s and refinement takes $4.4$\,s, for a total of $6.5$\,s per candidate.

We evaluate generation quality, geometric and mechanical conditioning, and the effects of the representation and refinement.
The implicit field of the structure is on a $150^3$ grid.
We extract the zero set using Marching Cubes
and compute curvature analytically at the extracted vertices.
The statistic $|H|_{\mathrm{avg}}$ is the average of per-sample absolute mean curvature.
Chamfer distance (CD) uses unsquared Euclidean distances,
while point-conditioning error is the mean prescribed-point-to-surface distance.
Detailed definitions are in Appendix~\ref{sec:evaluation_metrics}.
The common cell size makes the reported distances directly comparable across tasks.
Curvature is expressed in inverse-length units, so its interpretation also depends on this normalization.
We use the mean to summarize overall residual curvature.

\begin{table}[t]
    \centering
    \caption{\textbf{Generation-pipeline comparison.}
    Curvature ($|H|_{\mathrm{avg}}$) and maximum genus (``Max. genus'') are reported.
    Inversion CD is a separate reconstruction metric, scaled by $10^{-3}$.
    ``Periodic'' indicates whether periodicity is enforced by construction.
    ``Raw'' refers to the structure before the curvature refinement, 
    while ``Aft'' denotes the structure after the curvature refinement.
    }
    \label{tab:baseline_comparison}
    \resizebox{0.9\linewidth}{!}{
    \setlength{\tabcolsep}{4pt}
    \begin{tabular}{lrrrrrc}
        \toprule
        Method & Training (h) $\downarrow$  & Inversion CD $\downarrow$ & Raw $|H|_{\mathrm{avg}}$ $\downarrow$ & Aft $|H|_{\mathrm{avg}}$ $\downarrow$& Max. genus $\uparrow$  & Periodic\\
        \midrule
        Trigonometric & \textbf{0} & --- & 1.23 & 1.07 & 8 & \cmark\\
        SDF-Diffusion & 71 & 10.13 & 1.15 & 0.90& 12 & \xmark\\
        SDFusion      & 103 & 4.06 & 1.19 & 1.09& 15 & \xmark\\
        Standard Fourier Space & 35 & 3.94 & 2.28 & 1.40& \textbf{17} & \cmark\\
        \textbf{Ours} & 30 & \textbf{3.71} & \textbf{1.00} & \textbf{0.67}& \textbf{17} & \cmark\\
        \bottomrule
    \end{tabular}
    }
\end{table}

\subsection{Generation Quality and Baseline Comparisons}
\label{sec:random_gen}
\paragraph{Evaluation protocol.}
For each method, we generate $300$ candidates and retain the $150$ with the lowest $|H|_{\mathrm{avg}}$ for generation comparison.
Table~\ref{tab:baseline_comparison} therefore evaluates a \emph{design-screening} setting.
Our selected set has a mean curvature of $0.67$, whereas the average over all $300$ candidates is $1.16$.
We distinguish these populations because screening changes the evaluated distribution.
Inversion CD is measured separately by reconstructing existing examples from the dataset.

\paragraph{Baseline results.}
We compare our method with the widely used trigonometric baseline combining six standard implicit terms~\citep{yang2022high},
SDF-Diffusion~\citep{shim2023diffusion}, which learns directly from volumetric SDFs,
SDFusion~\citep{cheng2023sdfusion}, which uses a VQ-VAE representation before diffusion,
and a standard Fourier-space representation.
All generative baselines are trained on our dataset.
The results are reported in Table~\ref{tab:baseline_comparison}.
For each method, we report $|H|_{\mathrm{avg}}$ both before and after curvature refinement. 
For our method and the trigonometric baseline, 
we apply our Fourier-space refinement, 
whereas for the rest methods, 
we use the widely adopted mean-curvature flow (MCF)~\citep{desbrun1999implicit}.
Whether with or without curvature refinement, 
our pipeline achieves the lowest mean curvature and the largest maximum genus,
while consistently producing single-component structures.
The training-free trigonometric baseline exhibits approximately $1.6\times$ higher residual mean curvature
and lower structural diversity, as reflected by its maximum genus.
Histogram statistics are shown in Figure~\ref{fig:trig}.
The other generative baselines require longer training times than our method 
while yielding higher mean curvature and lower structural diversity.
The two SDF baselines also exhibit periodic-boundary inconsistencies in our implementations.
These results show that our pipeline performs better on the reported metrics.

\paragraph{Geometric variation.}
Samples generated by our method also exhibit geometric novelty, 
with an average CD of $9.1\times10^{-3}$ to the nearest training sample,
approximately $2.5\times$ the average inversion CD.
A selected set of visual results is provided in the Appendix,
in which Figure~\ref{fig:quantitative-gen} reports the distribution,
Figure~\ref{fig:random-gen-cd} compares candidates with their nearest dataset neighbors,
and Figure~\ref{fig:gallery} shows additional samples.
These examples demonstrate geometric variation beyond 
direct reproduction of the illustrated neighbors.

\begin{figure}
    \centering
    \includegraphics[width=1\linewidth]{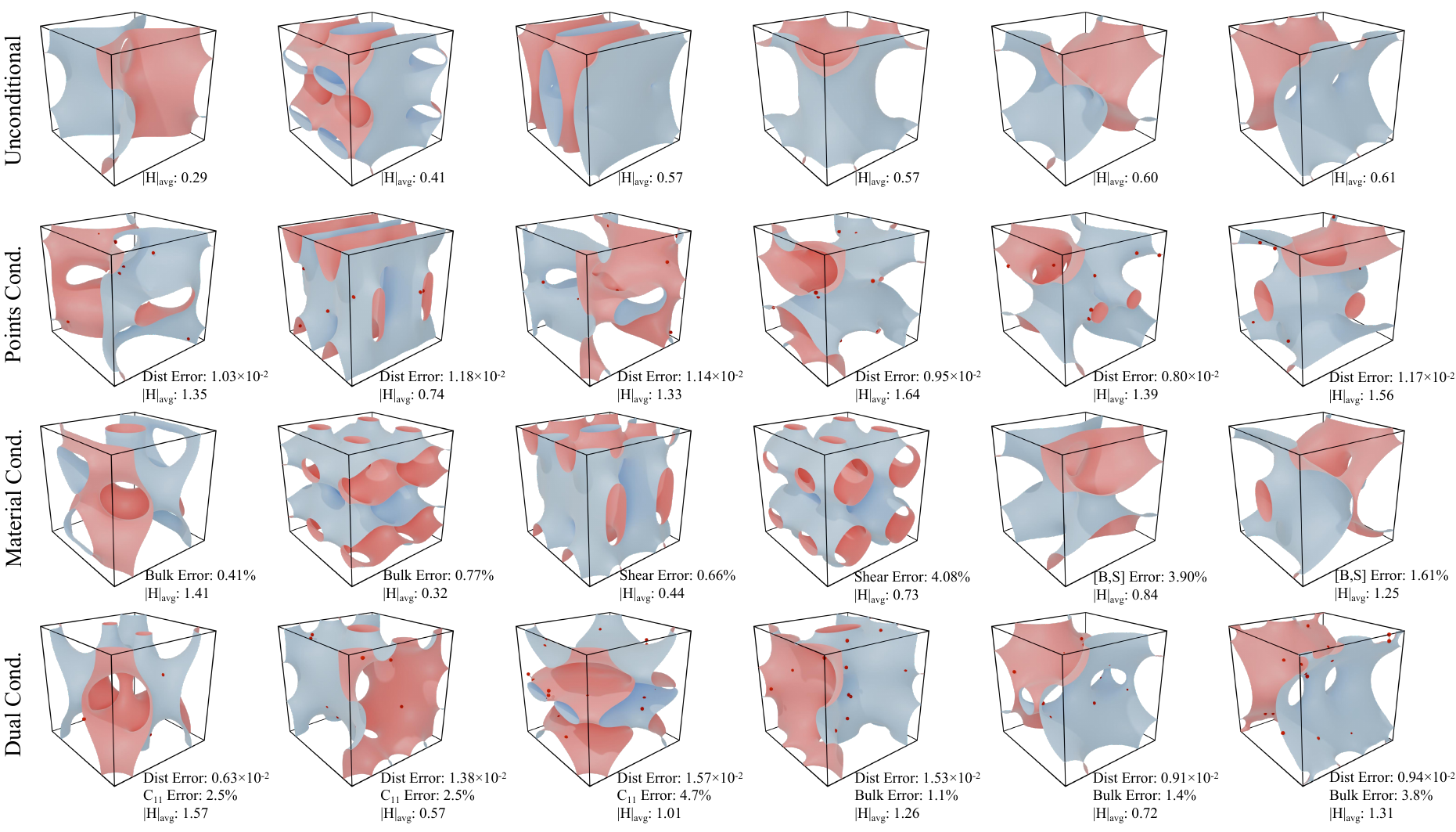}
    \vspace{-0.4cm}
    \caption{\textbf{Representative generation results under different conditioning settings.}
    From top to bottom: unconditional generation, point conditioning, material-property conditioning, and joint point--property conditioning. For material conditioning, the columns correspond to bulk, shear, and joint bulk--shear conditions; for joint conditioning, they correspond to point constraints combined with either $C_{11}$ or bulk modulus.}
    \label{fig:gen}
\end{figure}

\subsection{Geometric and Mechanical Conditioning}
\label{sec:conditioning}

\paragraph{Evaluation protocol.}
We evaluate sparse-point conditions, material-property conditions, and their combinations (Table~\ref{tab:conditioning}).
Each point condition consists of $16$ 3D points specified by their coordinates within the cell.
For mechanical evaluation, surfaces are uniformly thickened to $0.02$ and assigned an isotropic constituent with Young's modulus $E_s=1$ and Poisson's ratio $\nu_s=0.3$.
The same periodic homogenization solver provides training labels and evaluates generated structures~\citep{zhang2023optimized}.
Bulk and shear values are Voigt--Reuss--Hill scalar averages of the orthotropic stiffness tensor (Appendix~\ref{sec:elastic_properties}).
For a nonzero scalar target $m^*$, the relative error is $100|\hat m-m^*|/|m^*|$.
The selected properties capture complementary responses: bulk and shear summarize volumetric and shear behavior, while $C_{11}$ specifies an individual stiffness component.
They provide low-dimensional targets for evaluating controllability without requiring every independent tensor entry to be prescribed.

\paragraph{Single-modality conditions.}
Across $30$ point-conditioned samples,
the mean point-to-surface distance is $0.017$, or $1.7\%$ of the cell period, with mean curvature $1.28$.
For bulk, shear, and joint bulk--shear conditioning, we evaluate $20$ samples per setting.
Bulk-only conditioning yields a $2.8\%$ mean relative error, compared with $9.2\%$ for shear;
both retain low residual curvature in the reported evaluation.
Simultaneously conditioning on bulk and shear lead to errors of $6.2\%$ and $5.7\%$, respectively.
Figure~\ref{fig:gen} shows representative constraints and generated geometries.

\paragraph{Joint geometric and mechanical conditions.}
Point constraints control local surface placement,
whereas homogenized properties depend on the global geometry.
Combining points with either $C_{11}$ or bulk modulus tests whether the model can satisfy local spatial requirements and an effective mechanical response simultaneously.
Our method is capable of doing so.
The respective property errors are $9.3\%$ and $6.9\%$, 
with point distances of $0.019$ and $0.021$ across $20$ samples per setting,
as reported in Table~\ref{tab:conditioning}.
These results support simultaneous, approximate control within the selected design family.
Figure~\ref{fig:gen} shows representative constraints and generated geometries.

\begin{table}[t]
    \centering
    \caption{\textbf{Conditional-design results.}
    Distances are reported for a unit-period cell; property errors are percentages.
    ``---'' denotes an inapplicable or unreported quantity, rather than zero error.}
    \label{tab:conditioning}
    \resizebox{0.55\linewidth}{!}{
    \setlength{\tabcolsep}{6pt}
    \begin{tabular}{lrrrr}
        \toprule
        Condition & Samples & Avg. Dist & Prop. err (\%) & $|H|_{\mathrm{avg}}$\\
        \midrule
        Points & 30 & 0.017 & --- & 1.28\\
        Bulk modulus & 20 & --- & 2.8 & 0.91\\
        Shear modulus & 20 & --- & 9.2 & 1.21\\
        Bulk + shear & 20 & --- & 6.2, 5.7 & 1.09\\
        Points + $C_{11}$ & 20 & 0.019 & 9.3 & 1.30\\
        Points + bulk & 20 & 0.021 & 6.9 & 1.19\\
        \bottomrule
    \end{tabular}
    }
\end{table}

\subsection{Inversion and Editing}
\label{sec:inversion_summary}
To assess reconstruction, we invert and decode $2{,}000$ training shapes and separately evaluate validation examples.
The mean CDs are $3.71\times10^{-3}$ and $3.82\times10^{-3}$, with corresponding mean curvature values of $1.43$ and $1.47$.
Similar errors across the two splits support accurate reconstruction of held-out inputs,
complementing rather than replacing the generation evaluation.
Appendix~\ref{sec:inversion} provides additional details and the deterministic update equations,
while Figure~\ref{fig:shape-inv} provides visual comparisons.
Our model also supports TPMS editing:
users specify target curves on selected cell faces to guide modifications of an existing candidate.
As shown in Figure~\ref{fig:teaser}, the edited shapes match the prescribed boundary curves while retaining low mean curvature
and remaining close to the original geometry.

\subsection{Ablation Study}
Although the raw diffusion outputs already exhibit low residual mean curvature,
curvature refinement further reduces the average $|H|_{\mathrm{avg}}$ from $1.00$ to $0.67$.
Meanwhile, the genus remains unchanged, and the average CD to the original generated shape is only $1.60\times10^{-3}$,
confirming that the optimization acts primarily as a local refinement,
while the generative model plays the central role in producing low-mean-curvature structures.
Additionally, we find that neither increasing $K_{\mathrm{max}}$ nor
using a standard Fourier representation improves generation quality.
Moreover, performing inverse homogenization directly in our compact latent space
fails to reliably identify structures that satisfy user requirements while maintaining low residual mean curvature.
These results are reported in Appendix~\ref{sec:ablation-app}.

\section{Conclusion}
We presented a generative-model-based framework for controllable generation of TPMS
with low residual mean curvature.
Through qualitative and quantitative experiments, 
we demonstrated that the proposed model supports 
deterministic shape inversion, 
unconditional generation, and conditional generation under sparse geometric and material-property constraints.
The generated TPMS exhibit low mean curvature and compare favorably with existing baselines.
These results show that our framework provides a practical approach for TPMS generation and inverse design.
Several limitations remain. First, although the generated surfaces have low residual mean curvature, 
they are not guaranteed to be exactly minimal. 
Second, the current latent representation is restricted to \(D_{2h}\)-symmetric TPMS, 
which excludes broader classes of TPMS.
Failure cases are discussed in Appendix~\ref{sec:failure}.

\newpage
\subsection*{AI use statement}
In this work, we used OpenAI’s ChatGPT (GPT-5.6, GPT-6) and Codex to assist in polishing the language and presentation of the manuscript, and improving the layout and presentation of figures. These tools were not involved in developing the research idea, formulating the proposed methodology, designing or selecting models, generating experimental data, conducting experiments, or interpreting the results. All AI-assisted text, including literature descriptions and citations, was manually reviewed and verified against the original sources by the authors. We take full responsibility for the final content of this work, including all text, claims, results, and artifacts produced with the aid of generative AI.

\bibliography{iclr2027_conference}

@inproceedings{yang2019pointflow,
  title={Pointflow: 3d point cloud generation with continuous normalizing flows},
  author={Yang, Guandao and Huang, Xun and Hao, Zekun and Liu, Ming-Yu and Belongie, Serge and Hariharan, Bharath},
  booktitle={Proceedings of the IEEE/CVF international conference on computer vision},
  pages={4541--4550},
  year={2019}
}

@inproceedings{zheng2022sdf,
  title={SDF-StyleGAN: implicit SDF-based StyleGAN for 3D shape generation},
  author={Zheng, Xinyang and Liu, Yang and Wang, Pengshuai and Tong, Xin},
  booktitle={Computer Graphics Forum},
  volume={41},
  number={5},
  pages={52--63},
  year={2022},
  organization={Wiley Online Library}
}

@inproceedings{chou2023diffusion,
  title={Diffusion-sdf: Conditional generative modeling of signed distance functions},
  author={Chou, Gene and Bahat, Yuval and Heide, Felix},
  booktitle={Proceedings of the IEEE/CVF international conference on computer vision},
  pages={2262--2272},
  year={2023}
}

@article{ho2020denoising,
  title={Denoising diffusion probabilistic models},
  author={Ho, Jonathan and Jain, Ajay and Abbeel, Pieter},
  journal={Advances in neural information processing systems},
  volume={33},
  pages={6840--6851},
  year={2020}
}

@article{fan2025mesh,
  title={A Mesh Is Worth 512 Numbers: Spectral-domain Diffusion Modeling for High-dimension Shape Generation},
  author={Fan, Jiajie and Trigui, Amal and Bonfanti, Andrea and Dietrich, Felix and B{\"a}ck, Thomas and Wang, Hao},
  journal={arXiv preprint arXiv:2503.06485},
  year={2025}
}

@article{hu2024neural,
  title={Neural wavelet-domain diffusion for 3d shape generation, inversion, and manipulation},
  author={Hu, Jingyu and Hui, Ka-Hei and Liu, Zhengzhe and Li, Ruihui and Fu, Chi-Wing},
  journal={ACM transactions on graphics},
  volume={43},
  number={2},
  pages={1--18},
  year={2024},
  publisher={ACM New York, NY, USA}
}

@article{ning2024dctdiff,
  title={Dctdiff: Intriguing properties of image generative modeling in the dct space},
  author={Ning, Mang and Li, Mingxiao and Su, Jianlin and Jia, Haozhe and Liu, Lanmiao and Bene{\v{s}}, Martin and Chen, Wenshuo and Salah, Albert Ali and Ertugrul, Itir Onal},
  journal={arXiv preprint arXiv:2412.15032},
  year={2024}
}

@article{liu2023meshdiffusion,
  title={Meshdiffusion: Score-based generative 3d mesh modeling},
  author={Liu, Zhen and Feng, Yao and Black, Michael J and Nowrouzezahrai, Derek and Paull, Liam and Liu, Weiyang},
  journal={arXiv preprint arXiv:2303.08133},
  year={2023}
}

@inproceedings{cheng2023sdfusion,
  title={Sdfusion: Multimodal 3d shape completion, reconstruction, and generation},
  author={Cheng, Yen-Chi and Lee, Hsin-Ying and Tulyakov, Sergey and Schwing, Alexander G and Gui, Liang-Yan},
  booktitle={Proceedings of the IEEE/CVF conference on computer vision and pattern recognition},
  pages={4456--4465},
  year={2023}
}

@article{raissi2019physics,
  title={Physics-informed neural networks: A deep learning framework for solving forward and inverse problems involving nonlinear partial differential equations},
  author={Raissi, Maziar and Perdikaris, Paris and Karniadakis, George E},
  journal={Journal of Computational physics},
  volume={378},
  pages={686--707},
  year={2019},
  publisher={Elsevier}
}

@article{berzins2024geometry,
  title={Geometry-informed neural networks},
  author={Berzins, Arturs and Radler, Andreas and Volkmann, Eric and Sanokowski, Sebastian and Hochreiter, Sepp and Brandstetter, Johannes},
  journal={arXiv preprint arXiv:2402.14009},
  year={2024}
}

@article{li2020fourier,
  title={Fourier neural operator for parametric partial differential equations},
  author={Li, Zongyi and Kovachki, Nikola and Azizzadenesheli, Kamyar and Liu, Burigede and Bhattacharya, Kaushik and Stuart, Andrew and Anandkumar, Anima},
  journal={arXiv preprint arXiv:2010.08895},
  year={2020}
}

@article{li2023geometry,
  title={Geometry-informed neural operator for large-scale 3d pdes},
  author={Li, Zongyi and Kovachki, Nikola and Choy, Chris and Li, Boyi and Kossaifi, Jean and Otta, Shourya and Nabian, Mohammad Amin and Stadler, Maximilian and Hundt, Christian and Azizzadenesheli, Kamyar and others},
  journal={Advances in Neural Information Processing Systems},
  volume={36},
  pages={35836--35854},
  year={2023}
}

@article{wu2024transolver,
  title={Transolver: A fast transformer solver for pdes on general geometries},
  author={Wu, Haixu and Luo, Huakun and Wang, Haowen and Wang, Jianmin and Long, Mingsheng},
  journal={arXiv preprint arXiv:2402.02366},
  year={2024}
}

@article{chung2022diffusion,
  title={Diffusion posterior sampling for general noisy inverse problems},
  author={Chung, Hyungjin and Kim, Jeongsol and Mccann, Michael T and Klasky, Marc L and Ye, Jong Chul},
  journal={arXiv preprint arXiv:2209.14687},
  year={2022}
}

@inproceedings{baldan2026physics,
  title={Physics vs distributions: Pareto optimal flow matching with physics constraints},
  author={Baldan, Giacomo and Liu, Qiang and Guardone, Alberto and Thuerey, Nils},
  booktitle={International Conference on Learning Representations},
  volume={2026},
  pages={101215--101243},
  year={2026}
}

@inproceedings{xue2025mind,
  title={MIND: Microstructure inverse design with generative hybrid neural representation},
  author={Xue, Tianyang and Liu, Longdu and Lu, Lin and Henderson, Paul and Tang, Pengbin and Li, Haochen and Liu, Jikai and Zhao, Haisen and Peng, Hao and Bickel, Bernd},
  booktitle={Proceedings of the Special Interest Group on Computer Graphics and Interactive Techniques Conference Conference Papers},
  pages={1--12},
  year={2025}
}

@article{yang2026guided,
  title={Guided diffusion for fast inverse design of voxel-based mechanical metamaterials},
  author={Yang, Yanyan and Wang, Lili and Zhai, Xiaoya and Chen, Kai and Wu, Wenming and Zhao, Yunkai and Chen, Falai and Liu, Ligang and Fu, Xiao-Ming},
  journal={Smart Materials in Manufacturing},
  volume={4},
  pages={100129},
  year={2026},
  publisher={Elsevier}
}

@inproceedings{reitebuch2019discrete,
  title={Discrete gyroid surface},
  author={Reitebuch, Ulrich and Skrodzki, Martin and Polthier, Konrad},
  booktitle={Proceedings of bridges 2019: Mathematics, art, music, architecture, education, culture},
  pages={461--464},
  year={2019}
}

@article{al2021mslattice,
  title={MSLattice: A free software for generating uniform and graded lattices based on triply periodic minimal surfaces},
  author={Al-Ketan, Oraib and Abu Al-Rub, Rashid K},
  journal={Material Design \& Processing Communications},
  volume={3},
  number={6},
  pages={e205},
  year={2021},
  publisher={Wiley Online Library}
}

@article{jones2021tpms,
  title={TPMS designer: A tool for generating and analyzing triply periodic minimal surfaces},
  author={Jones, Alistair and Leary, Martin and Bateman, Stuart and Easton, Mark},
  journal={Software Impacts},
  volume={10},
  pages={100167},
  year={2021},
  publisher={Elsevier}
}

@article{maskery2022flatt,
  title={FLatt Pack: A research-focussed lattice design program},
  author={Maskery, Ian and Parry, Luke A and Padr{\~a}o, Daniel and Hague, Richard JM and Ashcroft, Ian A},
  journal={Additive Manufacturing},
  volume={49},
  pages={102510},
  year={2022},
  publisher={Elsevier}
}

@article{feng2021isotropic,
  title={Isotropic porous structure design methods based on triply periodic minimal surfaces},
  author={Feng, Jiawei and Liu, Bo and Lin, Zhiwei and Fu, Jianzhong},
  journal={Materials \& Design},
  volume={210},
  pages={110050},
  year={2021},
  publisher={Elsevier}
}

@article{khaleghi2021directional,
  title={On the directional elastic modulus of the TPMS structures and a novel hybridization method to control anisotropy},
  author={Khaleghi, Saeed and Dehnavi, Fayyaz N and Baghani, Mostafa and Safdari, Masoud and Wang, Kui and Baniassadi, Majid},
  journal={Materials \& Design},
  volume={210},
  pages={110074},
  year={2021},
  publisher={Elsevier}
}

@article{makatura2023procedural,
  title={Procedural metamaterials: A unified procedural graph for metamaterial design},
  author={Makatura, Liane and Wang, Bohan and Chen, Yi-Lu and Deng, Bolei and Wojtan, Chris and Bickel, Bernd and Matusik, Wojciech},
  journal={ACM Transactions on Graphics},
  volume={42},
  number={5},
  pages={1--19},
  year={2023},
  publisher={ACM New York, NY}
}

@article{feng2022triply,
  title={Triply periodic minimal surface (TPMS) porous structures: from multi-scale design, precise additive manufacturing to multidisciplinary applications},
  author={Feng, Jiawei and Fu, Jianzhong and Yao, Xinhua and He, Yong},
  journal={International Journal of Extreme Manufacturing},
  volume={4},
  number={2},
  pages={022001},
  year={2022},
  publisher={IOP Publishing}
}

@article{meeks2011classical,
  title={The classical theory of minimal surfaces},
  author={Meeks III, William and P{\'e}rez, Joaqu{\'\i}n},
  journal={Bulletin of the American Mathematical Society},
  volume={48},
  number={3},
  pages={325--407},
  year={2011}
}

@article{hsieh2020minisurf,
  title={Minisurf--A minimal surface generator for finite element modeling and additive manufacturing},
  author={Hsieh, Meng-Ting and Valdevit, Lorenzo},
  journal={Software Impacts},
  volume={6},
  pages={100026},
  year={2020},
  publisher={Elsevier}
}

@article{akbari2022strut,
  title={Strut-based cellular to shellular funicular materials},
  author={Akbari, Mostafa and Mirabolghasemi, Armin and Bolhassani, Mohammad and Akbarzadeh, Abdolhamid and Akbarzadeh, Masoud},
  journal={Advanced Functional Materials},
  volume={32},
  number={14},
  pages={2109725},
  year={2022},
  publisher={Wiley Online Library}
}

@inproceedings{akbari2020geometry,
  title={Geometry-based structural form-finding to design architected cellular solids},
  author={Akbari, Mostafa and Mirabolghasemi, Armin and Akbarzadeh, Hamid and Akbarzadeh, Masoud},
  booktitle={Proceedings of the ACM Symposium on Computational Fabrication},
  pages={1--11},
  year={2020}
}

@article{pinkall1993computing,
  title={Computing discrete minimal surfaces and their conjugates},
  author={Pinkall, Ulrich and Polthier, Konrad},
  journal={Experimental mathematics},
  volume={2},
  number={1},
  pages={15--36},
  year={1993},
  publisher={Taylor \& Francis}
}

@article{wang2021computing,
  title={Computing minimal surfaces with differential forms},
  author={Wang, Stephanie and Chern, Albert},
  journal={ACM Transactions on Graphics (TOG)},
  volume={40},
  number={4},
  pages={1--14},
  year={2021},
  publisher={ACM New York, NY, USA}
}

@article{xu2023new,
  title={New families of triply periodic minimal surface-like shell lattices},
  author={Xu, Yonglai and Pan, Hao and Wang, Ruonan and Du, Qiang and Lu, Lin},
  journal={Additive Manufacturing},
  volume={77},
  pages={103779},
  year={2023},
  publisher={Elsevier}
}

@article{zhang2025asymptotic,
  title={Asymptotic analysis and design of linear elastic shell lattice metamaterials},
  author={Zhang, Di and Liu, Ligang},
  journal={ACM Transactions on Graphics (TOG)},
  volume={44},
  number={4},
  pages={1--18},
  year={2025},
  publisher={ACM New York, NY, USA}
}

@inproceedings{park2019deepsdf,
  title={Deepsdf: Learning continuous signed distance functions for shape representation},
  author={Park, Jeong Joon and Florence, Peter and Straub, Julian and Newcombe, Richard and Lovegrove, Steven},
  booktitle={Proceedings of the IEEE/CVF conference on computer vision and pattern recognition},
  pages={165--174},
  year={2019}
}

@inproceedings{ronneberger2015u,
  title={U-net: Convolutional networks for biomedical image segmentation},
  author={Ronneberger, Olaf and Fischer, Philipp and Brox, Thomas},
  booktitle={International Conference on Medical image computing and computer-assisted intervention},
  pages={234--241},
  year={2015},
  organization={Springer}
}

@article{song2020score,
  title={Score-based generative modeling through stochastic differential equations},
  author={Song, Yang and Sohl-Dickstein, Jascha and Kingma, Diederik P and Kumar, Abhishek and Ermon, Stefano and Poole, Ben},
  journal={arXiv preprint arXiv:2011.13456},
  year={2020}
}

@inproceedings{bao2023all,
  title={All are worth words: A vit backbone for diffusion models},
  author={Bao, Fan and Nie, Shen and Xue, Kaiwen and Cao, Yue and Li, Chongxuan and Su, Hang and Zhu, Jun},
  booktitle={Proceedings of the IEEE/CVF conference on computer vision and pattern recognition},
  pages={22669--22679},
  year={2023}
}

@article{song2020denoising,
  title={Denoising diffusion implicit models},
  author={Song, Jiaming and Meng, Chenlin and Ermon, Stefano},
  journal={arXiv preprint arXiv:2010.02502},
  year={2020}
}

@article{dhariwal2021diffusion,
  title={Diffusion models beat gans on image synthesis},
  author={Dhariwal, Prafulla and Nichol, Alexander},
  journal={Advances in neural information processing systems},
  volume={34},
  pages={8780--8794},
  year={2021}
}

@article{kingma2014adam,
  title={Adam: A method for stochastic optimization},
  author={Kingma, Diederik P and Ba, Jimmy},
  journal={arXiv preprint arXiv:1412.6980},
  year={2014}
}

@article{loshchilov2017decoupled,
  title={Decoupled weight decay regularization},
  author={Loshchilov, Ilya and Hutter, Frank},
  journal={arXiv preprint arXiv:1711.05101},
  year={2017}
}

@inproceedings{shim2023diffusion,
  title={Diffusion-based signed distance fields for 3d shape generation},
  author={Shim, Jaehyeok and Kang, Changwoo and Joo, Kyungdon},
  booktitle={Proceedings of the IEEE/CVF conference on computer vision and pattern recognition},
  pages={20887--20897},
  year={2023}
}

@article{yang2022high,
  title={High-throughput generation of 3D graphene metamaterials and property quantification using machine learning},
  author={Yang, Zhenze and Buehler, Markus J},
  journal={Small Methods},
  volume={6},
  number={9},
  pages={2200537},
  year={2022},
  publisher={Wiley Online Library}
}

@article{zhang2023optimized,
  title={An optimized, easy-to-use, open-source GPU solver for large-scale inverse homogenization problems},
  author={Zhang, Di and Zhai, Xiaoya and Liu, Ligang and Fu, Xiao-Ming},
  journal={Structural and Multidisciplinary Optimization},
  volume={66},
  number={9},
  pages={207},
  year={2023},
  publisher={Springer}
}

@inproceedings{desbrun1999implicit,
  title={Implicit fairing of irregular meshes using diffusion and curvature flow},
  author={Desbrun, Mathieu and Meyer, Mark and Schr{\"o}der, Peter and Barr, Alan H},
  booktitle={Proceedings of the 26th annual conference on Computer graphics and interactive techniques},
  pages={317--324},
  year={1999}
}

@article{torquato2003optimal,
  title={Optimal design of manufacturable three-dimensional composites with multifunctional characteristics},
  author={Torquato, S and Hyun, S and Donev, Aleksandar},
  journal={Journal of Applied Physics},
  volume={94},
  number={9},
  pages={5748--5755},
  year={2003},
  publisher={American Institute of Physics}
}
\bibliographystyle{iclr2027_conference}

\newpage
\appendix

\section{dataset details}
\label{sec:dataset_details}

\begin{figure}[h]
    \centering
    \includegraphics[width=0.8\linewidth]{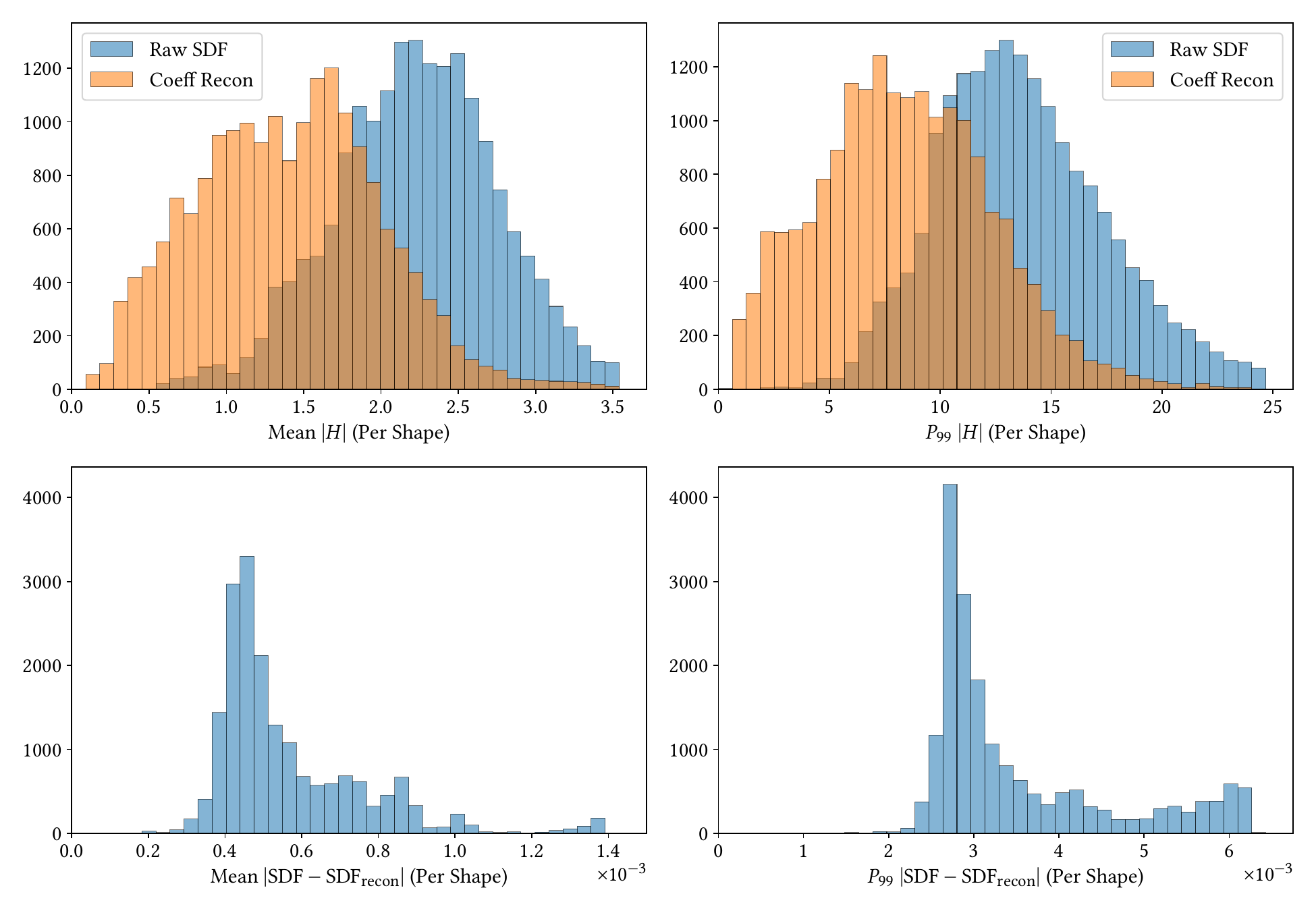}
    \caption{\textbf{Numerical results of data processing.} Top row: after our symmetry processing and frequency-domain truncation, the shapes exhibit a significant reduction in mean curvature, better satisfying the basic requirements of TPMS. Bottom row: meanwhile, we maintain a tiny difference between the original and processed shapes, with the average SDF difference remaining on the order of $10^{-4}$.}
    \label{fig:stats-dataset}
\end{figure}

\paragraph{Notation.}
Every shape lives in the periodic unit cell $\Omega=[0,1)^3$.
A field sampled on an $n^3$ grid is written $f_{\vp}$ with $\vp=(p_1,p_2,p_3)\in\{0,\dots,n-1\}^3$,
indices taken modulo $n$.
Its discrete Fourier transform is $\hat f_{\vq}=\sum_{\vp} f_{\vp}\,e^{-2\pi\mathrm{i}\,\vq\cdot\vp/n}$.
For $\vb\in\{0,1\}^3$, $\tau_{\vb}$ is the circular shift by half a period along every axis $a$ with $b_a=1$,
$(\tau_{\vb}f)_{\vp}=f_{\vp+\frac{n}{2}\vb}$.
$\bar f$ is the mean of $f$ over the grid, and $[\,\cdot\,]$ is $1$ if the condition holds and $0$ otherwise.

\begin{enumerate}[label=\textbf{Step~\arabic*.}, ref=\arabic*, leftmargin=*, align=left, itemsep=1.5ex]
\item \label{step:mesh}
\textbf{Mesh generation and curvature filtering.}
TPMS surface meshes are generated with the procedural graph of~\citet{makatura2023procedural}.
We use the conditions that check if a loop is valid from~\citet{makatura2023procedural} and combine the loop variations with eight different boundary types.
The mean curvature $H$ is computed on each mesh, and meshes whose average mean curvature exceeds $7.5$ are discarded.

\item \label{step:rotation}
\textbf{Rotation augmentation.}
Each mesh is rotated by the rotations of the cube that map the unit cell onto itself
(rotations by multiples of $90^\circ$ about the coordinate axes).
Only the distinct rotated copies are kept, and each is added to the dataset as a separate shape.

\item \label{step:sdf}
\textbf{Signed distance field.}
\begin{enumerate}[label=(\alph*)]
  \item Sample the cell on a $257\times257\times257$ grid of points $\vx_{\vp}=\vp/256$, $\vp\in\{0,\dots,256\}^3$;
        both boundary faces of the cell are included.
  \item Compute the unsigned distance $d_{\vp}$ from $\vx_{\vp}$ to the mesh.
  \item Assign signs by flood fill: the 8 corners of the cell are assumed to be positive, and the sign is
        propagated through the grid, giving the signed distance field $\phi_{\vp}=\pm d_{\vp}$.
  \item Keep the field only if the maximum distance error reported by the distance computation is below
        half a voxel, $0.5/256$.
  \item Because $\phi$ is periodic, the samples with $p_a=256$ repeat those with $p_a=0$.
        They are dropped, leaving the periodic $256^3$ grid $\phi_{\vp}=\phi(\vp/256)$, $\vp\in\{0,\dots,255\}^3$,
        which all later steps use.
\end{enumerate}

\item \label{step:align}
\textbf{DC removal, canonical alignment, and deduplication.}
The field is mirror-symmetric about the origin in $x$, $y$ and $z$, and by periodicity also about every point
of $\{0,\tfrac12\}^3$.
The only remaining freedom is therefore the global sign and which of these 8 symmetry centers sits at the origin,
which gives 16 equivalent fields $s\,\tau_{\vb}\phi$ with $s\in\{+1,-1\}$ and $\vb\in\{0,1\}^3$.
Their Fourier coefficients are $s\,(-1)^{\vq\cdot\vb}\,\hat\phi_{\vq}$: a half-period shift along $x$ multiplies
mode $(h,k,l)$ by $(-1)^h$, and a sign flip multiplies every mode by $-1$.
\begin{enumerate}[label=(\alph*)]
  \item Remove the DC component: $\phi\leftarrow\phi-\bar\phi$. 
  Note that since the original shape is already close to TPMS, $\bar{\phi}$ is close to zero.
  Thus, this adjustment does not change the represented surface much.
  \item Take every 32nd voxel, $\psi_{\vp}=\phi_{32\vp}$ for $\vp\in\{0,\dots,7\}^3$, and compute its $8^3$ DFT
        $\hat\psi$.
  \item Dominant mode: $\vq^\star=\arg\max_{\vq\neq\mathbf{0}}\,|\Re\hat\psi_{\vq}|$.
  \item Keep the 8 of the 16 versions in which this coefficient is positive,
        \[
          \mathcal{V}=\bigl\{(s,\vb):\ s\,(-1)^{\vq^\star\cdot\vb}\,\Re\hat\psi_{\vq^\star}>0\bigr\}.
        \]
  \item Tie-break: choose $(s^\star,\vb^\star)\in\mathcal{V}$ whose subsample $s\,\tau_{\vb}\psi$, flattened in
        row-major order, is lexicographically smallest; entries that differ by at most $10^{-12}$ count as equal.
        The first entry is the value at the origin, so in practice the version with the smallest $\phi$ at the
        origin wins.
  \item Apply the winning shift and sign at full resolution, $\phi^\star=s^\star\,\tau_{\vb^\star}\phi$ on the
        $256^3$ grid, and store it in single precision.
  \item Downsample $\phi^\star$ to $16^3$, $32^3$, and $64^3$. 
  We first compare each pair of SDF samples at the coarsest resolution. 
  If they are identical, we repeat the comparison at progressively finer resolutions up to $64^3$. 
  This procedure removes all duplicate samples.
\end{enumerate}

\item \label{step:curvature}
\textbf{Curvature filtering of the aligned field.}
\begin{enumerate}[label=(\alph*)]
  \item On the $256^3$ grid, with periodic central differences of spacing $1/256$, compute $H$ as described in Appendix~\ref{sec:fourier_to_curvature}.
  \item Extract the zero level set of $\phi^\star$ with marching cubes and sample 4{,}096 of its vertices uniformly
        at random.
        Move each sample onto the surface with one Newton step,
        $\vx\leftarrow\vx-\phi^\star(\vx)\,\nabla\phi^\star(\vx)/\|\nabla\phi^\star(\vx)\|^2$,
        and evaluate $H$ there by trilinear interpolation.
  \item Keep the shape if, over the samples, $\operatorname{mean}|H|<7.5$ and $\max|H|<100$.
\end{enumerate}
After this step, 18{,}704 shapes remain.

\end{enumerate}

\paragraph{Output.}
The final dataset has 18{,}704 shapes. Figure~\ref{fig:stats-dataset} shows the statistics.

\section{Basis computation details}
\label{sec:basis_details}
Let $r_x(\mathbf p)=(L-p_x,p_y,p_z)$, with $r_y$ and $r_z$ defined analogously.
These commuting reflections generate an eight-element group,
\begin{equation}
    G=\{r_x^\alpha r_y^\beta r_z^\gamma:\alpha,\beta,\gamma\in\{0,1\}\}
    \cong\mathbb Z_2^3\cong D_{2h}.
\end{equation}
The requirement is invariance of the signed field, $f(g\mathbf p)=f(\mathbf p)$, rather than only invariance of its unsigned zero set.
A general real periodic field has expansion
\begin{equation}
    f(\mathbf p)=\sum_{\mathbf n\in\mathbb Z^3}
    c_{\mathbf n}\exp\!\left(\frac{2\pi i}{L}\mathbf n\cdot\mathbf p\right),
    \qquad c_{-\mathbf n}=\overline{c_{\mathbf n}}.
\end{equation}
Coordinate-reflection invariance ties coefficients under independent sign reversals of the three frequency indices.
Grouping these modes yields the cosine-product expansion in Eq.~(\ref{eq:d2h-basis}).
This derivation explains the reduction: 
the model need not learn repeated coefficients corresponding to the same symmetry orbit.
It does not add information beyond the corresponding symmetry-constrained Fourier subspace.

For normalized DFT coefficients $F_{hk\ell}$ of the symmetrized field, define
$q(h,k,\ell)=\mathbf 1_{h>0}+\mathbf 1_{k>0}+\mathbf 1_{\ell>0}$.
Away from Nyquist-index special cases, the conversion to cosine-product coefficients is
\begin{equation}
    a_{hk\ell}=2^{q(h,k,\ell)}\operatorname{Re}F_{hk\ell}.
    \label{eq:coefficient_projection}
\end{equation}
Each nonzero cosine factor contributes two exponential modes with weights $1/2$.
The constant coefficient is set to zero as described above.

The count $(2K_{\max}+1)^3$ for the general real Fourier field already accounts for conjugate symmetry; it counts independent real scalars, not independently stored complex numbers.
The invariant basis has $(K_{\max}+1)^3$ real coefficients, a reduction approaching a factor of eight as $K_{\max}$ increases.
At $K_{\max}=15$, the ratio is approximately $7.27$.
The $4{,}096$ entries in our $16^3$ tensor include the constant component; fixing that component leaves $4{,}095$ adjustable entries if the same restriction is maintained during generation.

\section{Implementation and Diffusion Details}
\label{sec:implementation}
\subsection{Normalization and Its Inverse}
For a positive scale parameter $\tau$, define the componentwise signed-logarithmic transform
\begin{equation}
    \phi_\tau(a)=\operatorname{sign}(a)\log\!\left(1+\frac{|a|}{\tau}\right),
    \qquad
    \phi_\tau^{-1}(u)=\tau\operatorname{sign}(u)\bigl(e^{|u|}-1\bigr).
    \label{eq:log_transform}
\end{equation}
In this paper, we set \(\tau=10^{-5}\). Let $\mathcal A$ denote the affine rescaling used to map the transformed training coefficients to $[-1,1]$, and let $\mathcal P$ denote patchification.
The encoding and decoding operations are
\begin{equation}
    \bm y=\mathcal T(\bm a)=\mathcal P\mathcal A\phi_\tau(\bm a),
    \qquad
    \bm a=\mathcal T^{-1}(\bm y)
    =\phi_\tau^{-1}\!\left(\mathcal A^{-1}\mathcal P^{-1}\bm y\right).
    \label{eq:normalization_inverse}
\end{equation}
Thus decoding first undoes patchification and affine rescaling, then applies the inverse logarithmic transform.
Neither transform imposes the Eikonal equation on the generated field.

\subsection{Neural Network Architecture}
\label{sec:neural-network}

\begin{figure}
    \centering
    \includegraphics[width=0.7\linewidth]{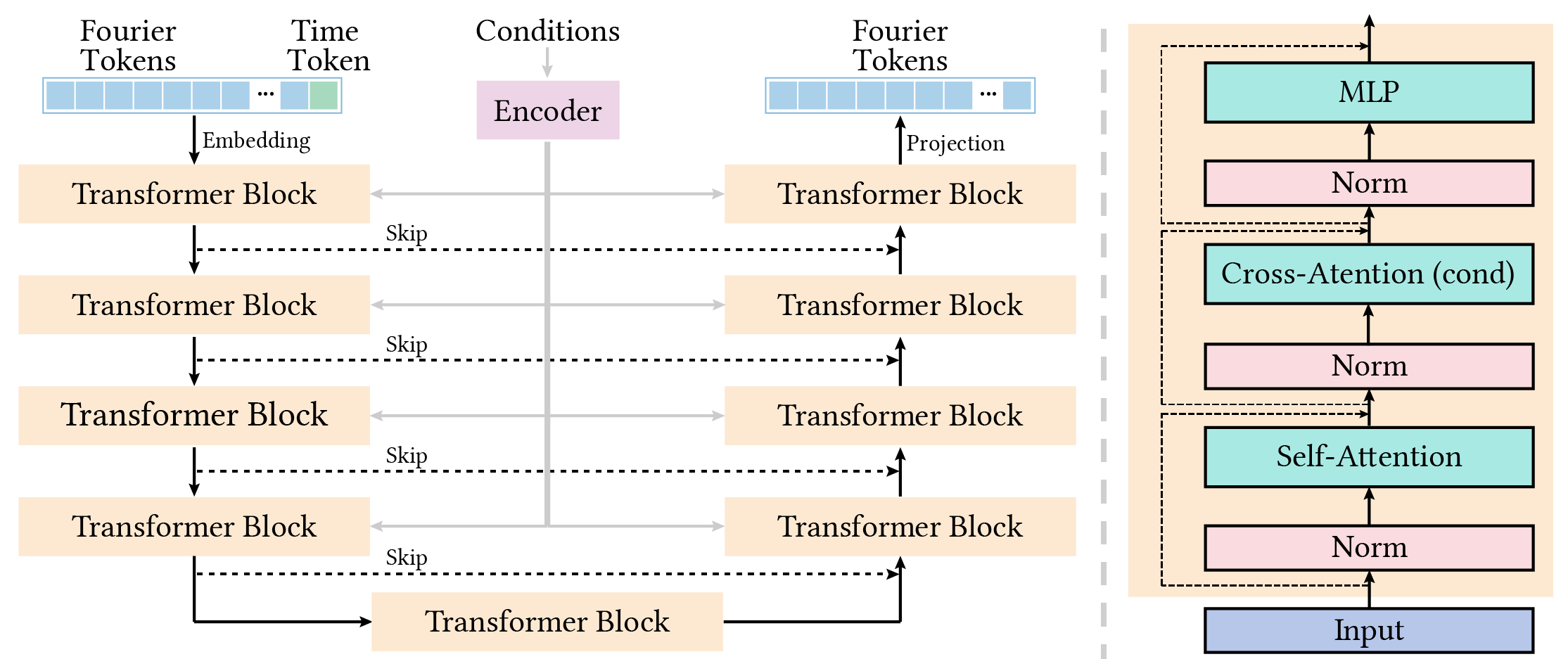}
    \vspace{-0.3cm}
    \caption{\textbf{Our neural network architecture.} 
    We use a transformer-based diffusion model with skip connections. 
    Geometric or material-property conditions are encoded as tokens 
    and injected into each transformer block through cross-attention.}
    \vspace{-0.5cm}
    \label{fig:network-arch}
\end{figure}

The network follows a U-shaped transformer design consisting of encoder blocks, a bottleneck block, and decoder blocks with skip fusion.
A final linear prediction head maps the decoded features back to the original token dimension, as shown in Figure~\ref{fig:network-arch}.
In our experiments, we use an embedding dimension of $D=256$, four encoder blocks, one bottleneck block, four decoder blocks, 
and four attention heads.

\subsection{Forward Schedule}
The variance-preserving forward process is
\begin{equation}
    d\bm y=-\tfrac12\beta(t)\bm y\,dt+\sqrt{\beta(t)}\,d\mathbf w,
    \qquad t\in[0,1],
\end{equation}
where $\mathbf w$ is a Wiener process.
Its perturbation coefficients in Eq.~(\ref{eq:diffusion_training}) are
\begin{equation}
    \alpha(t)=\exp\!\left(-\tfrac12\int_0^t\beta(s)\,ds\right),
    \qquad \sigma(t)=\sqrt{1-\alpha(t)^2}.
    \label{eq:vp_schedule}
\end{equation}
The reported linear schedule uses $\beta_{\min}=0.1$ and $\beta_{\max}=10.0$.
The denoiser predicts the Gaussian perturbation, using the squared-error objective in Eq.~(\ref{eq:diffusion_training}).

\section{Inversion Details}
\label{sec:inversion}

\begin{figure}[h]
    \centering
    \includegraphics[width=0.8\linewidth]{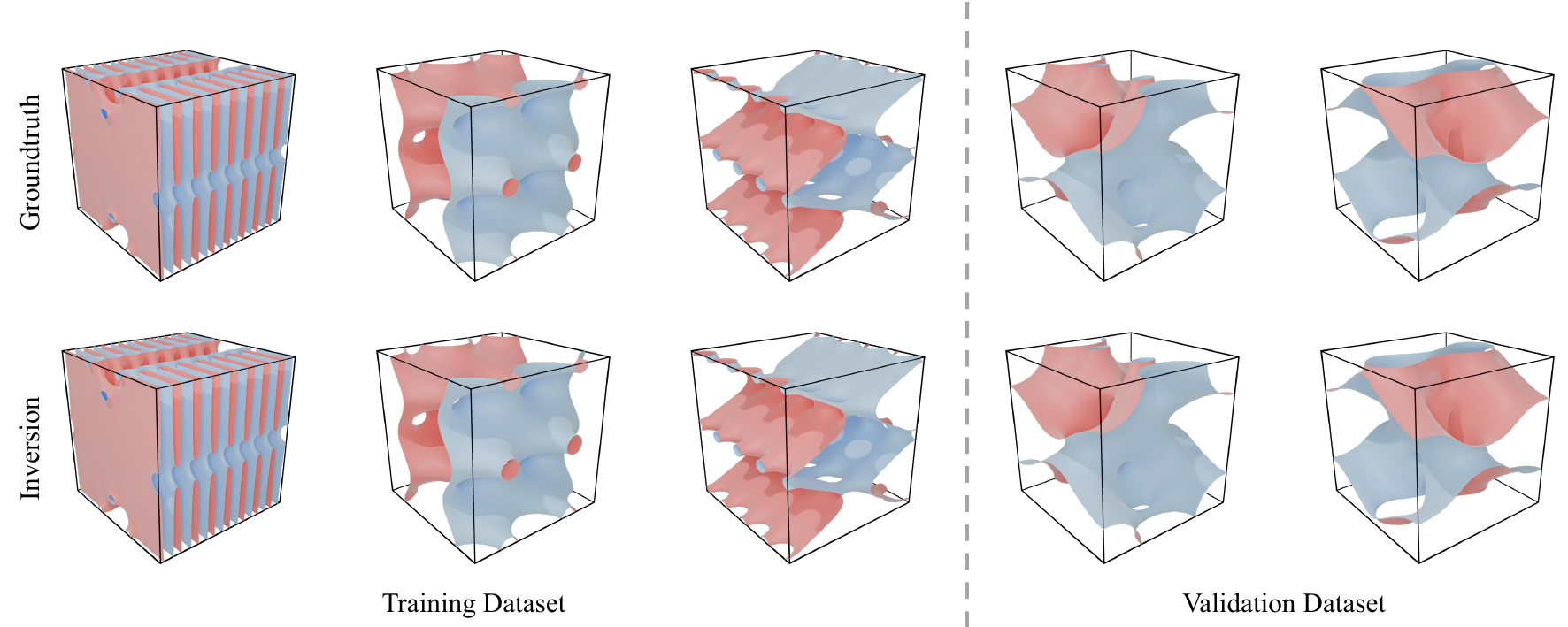}
    \caption{\textbf{Visualization of deterministic shape inversion.} Our model achieves high-fidelity reconstruction for both seen and unseen shapes, showing that the learned diffusion trajectory is stable and generalizable.}
    \label{fig:shape-inv}
\end{figure}

\begin{figure}[h]
    \centering
    \includegraphics[width=0.8\linewidth]{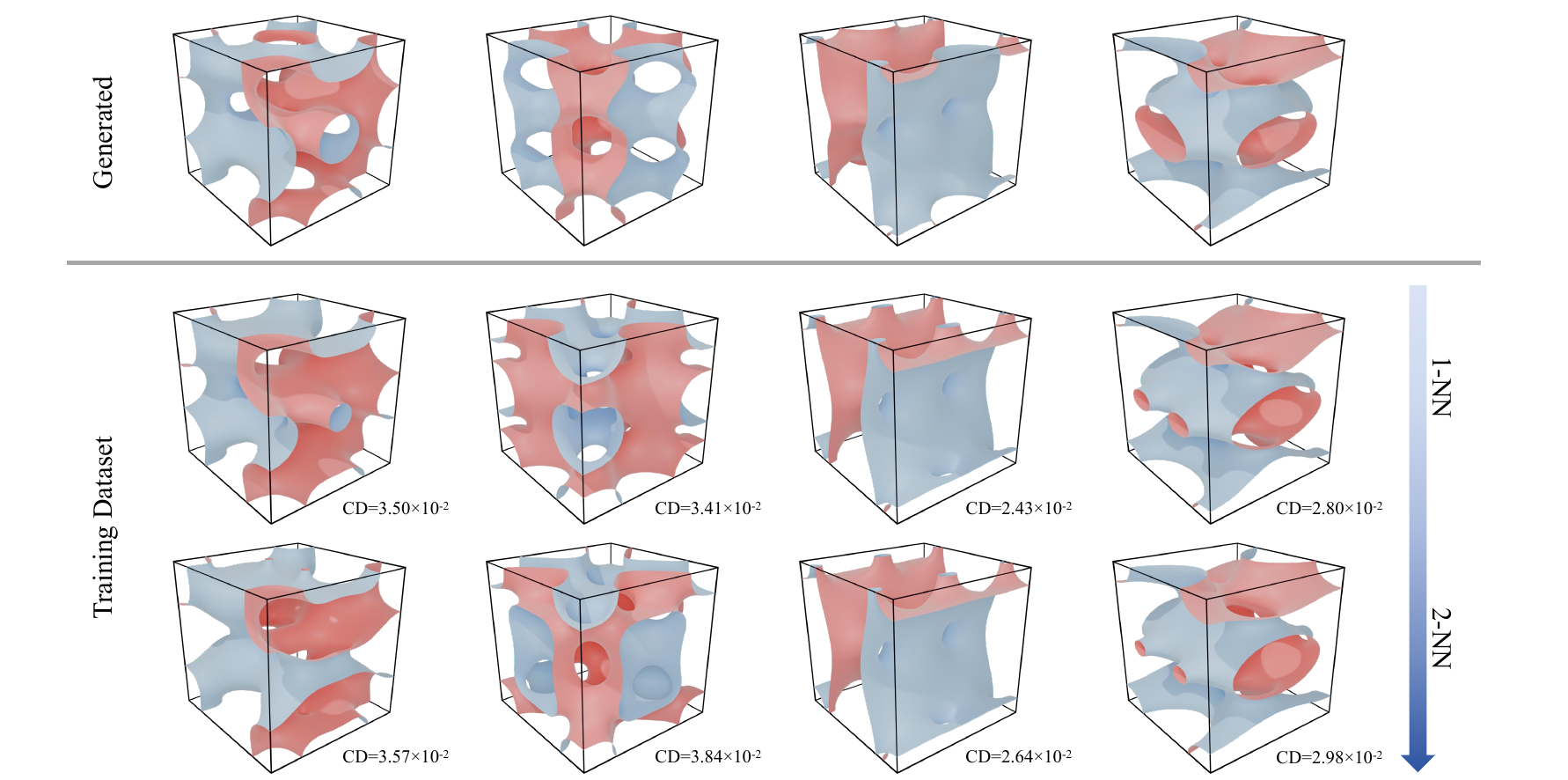}
    \caption{\textbf{Shape novelty analysis.} First row: randomly generated shapes; second and third rows: the shapes from the training set with the smallest and second-smallest CD, respectively. Our generated results exhibit clear topological differences from the training data.}
    \label{fig:random-gen-cd}
\end{figure}

\subsection{Deterministic Updates}
At time $t_i$, the denoiser predicts a noise tensor and a clean-token estimate,
\begin{equation}
    \hat{\boldsymbol\epsilon}_i=\boldsymbol\epsilon_\theta(\bm y_{t_i},t_i,\mathbf c),
    \qquad
    \hat{\bm y}_0=\frac{\bm y_{t_i}-\sigma(t_i)\hat{\boldsymbol\epsilon}_i}{\alpha(t_i)}.
    \label{eq:clean_prediction}
\end{equation}
The deterministic DDIM-style update is
\begin{equation}
    \bm y_{t_j}=\alpha(t_j)\hat{\bm y}_0+\sigma(t_j)\hat{\boldsymbol\epsilon}_i.
    \label{eq:ddim_update}
\end{equation}
Here $\hat{\boldsymbol\epsilon}_i$ is a network prediction, not an independent random draw.
For a fixed initial state, condition, and timestep sequence, the update is deterministic.
Generation starts from a Gaussian state and uses decreasing noise levels, $t_j<t_i$.
Inversion starts from encoded geometry and uses increasing levels, 
$t_j>t_i$; traversing the reverse sequence then approximately reconstructs the input.

\subsection{Guided Editing}
For local editing, the predicted clean coefficients are
$\bm a_{\mathrm{pred}}=\mathcal T^{-1}(\hat{\bm y}_0)$.
Following gradient-guided inference~\citep{dhariwal2021diffusion,chung2022diffusion}, 
we evaluate a differentiable loss on this prediction.
For prescribed points $\mathcal P_{\mathrm{con}}$, an implicit-field penalty is
\begin{equation}
    \mathcal L_{\mathrm{geom}}(\hat{\bm y}_0;\mathcal P_{\mathrm{con}})=
    \frac{1}{|\mathcal P_{\mathrm{con}}|}
    \sum_{\mathbf p\in\mathcal P_{\mathrm{con}}}
    |f(\mathbf p;\mathcal T^{-1}(\hat{\bm y}_0))|^2.
    \label{eq:editing_loss}
\end{equation}
The guided estimate is
\begin{equation}
    \tilde{\bm y}_0=\hat{\bm y}_0-
    \eta_t\nabla_{\hat{\bm y}_0}\mathcal L_{\mathrm{guide}}(\hat{\bm y}_0;\mathbf c_{\mathrm{edit}}),
    \label{eq:editing_update}
\end{equation}
where $\eta_t$ controls guidance strength and $\mathcal L_{\mathrm{guide}}$ may be the geometric penalty above.
We substitute $\tilde{\bm y}_0$ for $\hat{\bm y}_0$ in Eq.~(\ref{eq:ddim_update}).
This operation encourages the zero set to pass through the target points, but $|f|$ is not a Euclidean distance unless the field has the SDF property.
Geometric evaluation therefore uses the actual point-to-surface distance defined in Appendix~\ref{sec:evaluation_metrics}.
The editing examples in Figure~\ref{fig:teaser} use target curves on selected cell faces and retain the overall morphology in the illustrated cases.

\subsection{Reconstruction Evaluation}
We invert and reconstruct $2{,}000$ randomly selected training shapes and separately evaluate validation shapes.
Table~\ref{tab:inversion} reports the provided statistics.
Representative reconstructions appear in Figure~\ref{fig:shape-inv}.

\begin{table}[h]
    \centering
    \caption{\textbf{Inversion and reconstruction.}
    CD values use unsquared distances in a unit-period cell.
    The curvature columns average the per-surface mean and per-surface 99th percentile, respectively.}
    \label{tab:inversion}
    \small
    \begin{tabular}{lrrrr}
        \toprule
        Split & Mean CD & 99th-percentile CD & Mean $|H|_{\mathrm{avg}}$ & Mean $|H|_{\mathrm{p99}}$\\
        \midrule
        Training & $3.71\times10^{-3}$ & $9.56\times10^{-3}$ & 1.43 & 9.18\\
        Validation & $3.82\times10^{-3}$ & $1.11\times10^{-2}$ & 1.47 & 9.31\\
        \bottomrule
    \end{tabular}
\end{table}

\section{Additional Results}
\label{sec:additional-ret}

\begin{figure}[h]
    \centering
    \includegraphics[width=0.9\linewidth]{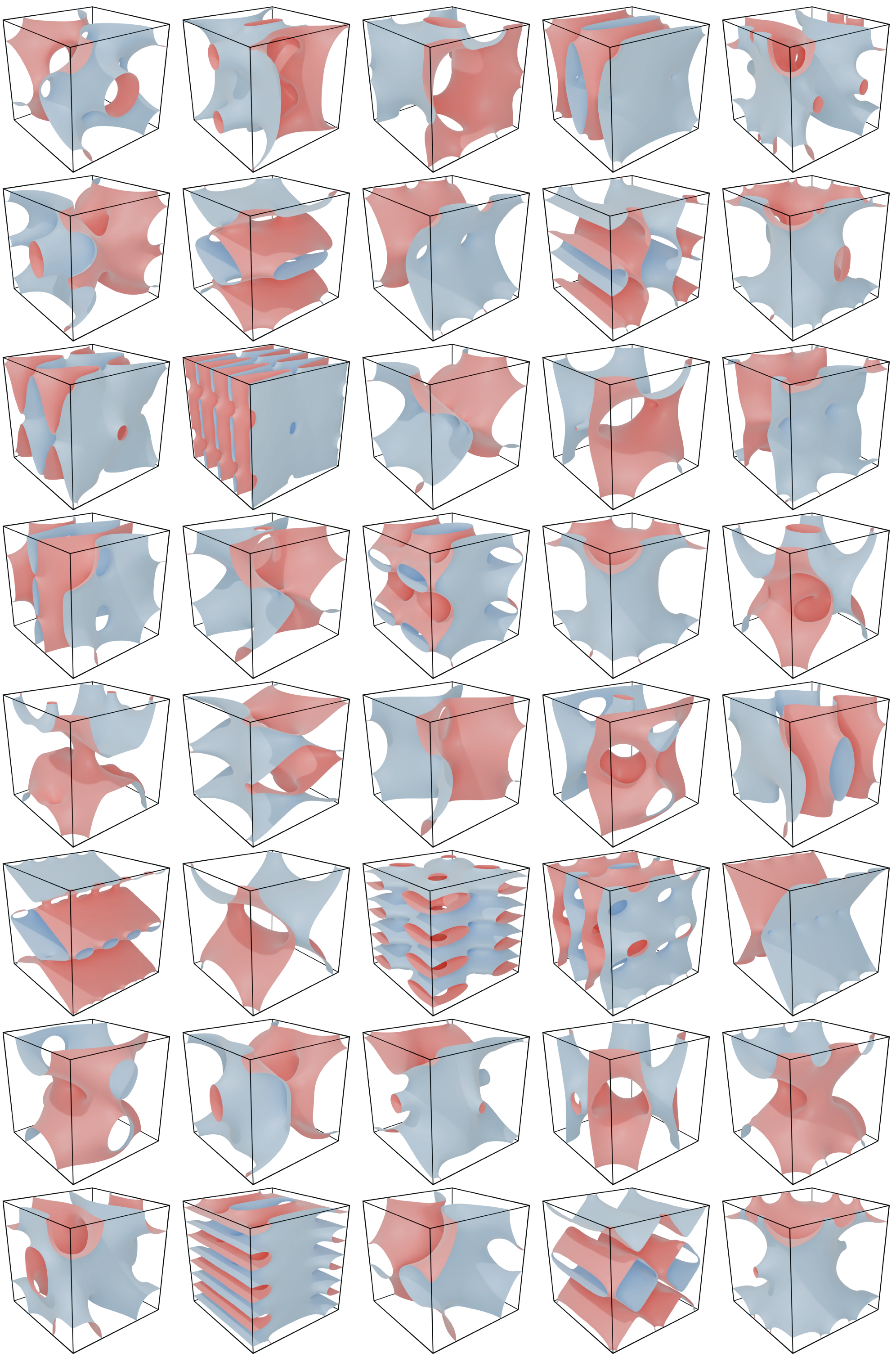}
    \caption{\textbf{Randomly generated TPMS.} Our model enables unconditional generation of TPMS from randomly initialized Gaussian noise. The generated shapes demonstrate rich geometric and topological diversity while preserving low mean curvature.}
    \label{fig:gallery} 
\end{figure}

\begin{figure}[h]
    \centering
    \begin{minipage}{0.4\columnwidth}
        \centering
        \includegraphics[width=\linewidth]{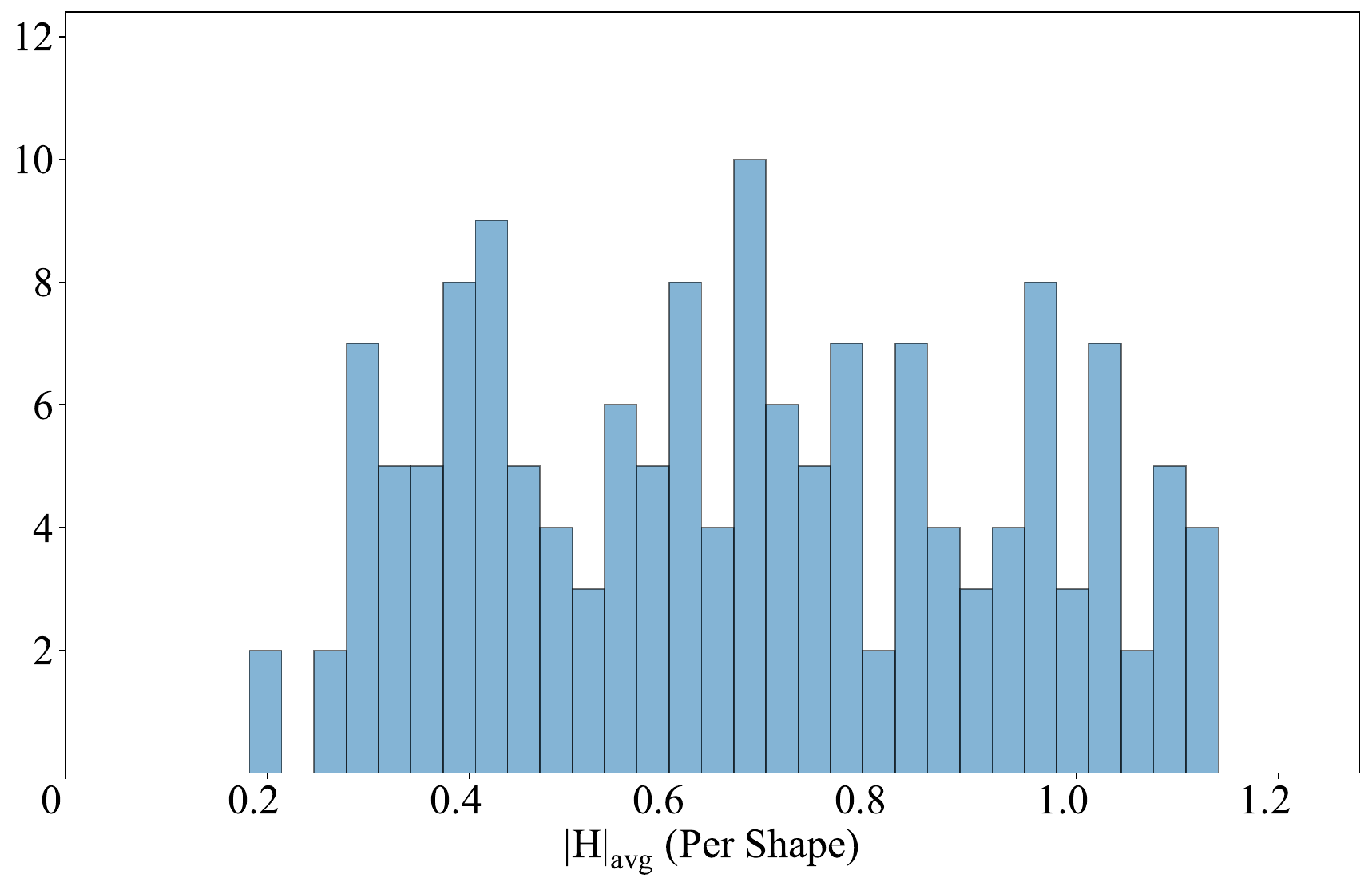}
        \label{generation_cham}
    \end{minipage}
    \hspace{-5pt}
    \begin{minipage}{0.4\columnwidth}
        \centering
        \includegraphics[width=\linewidth]{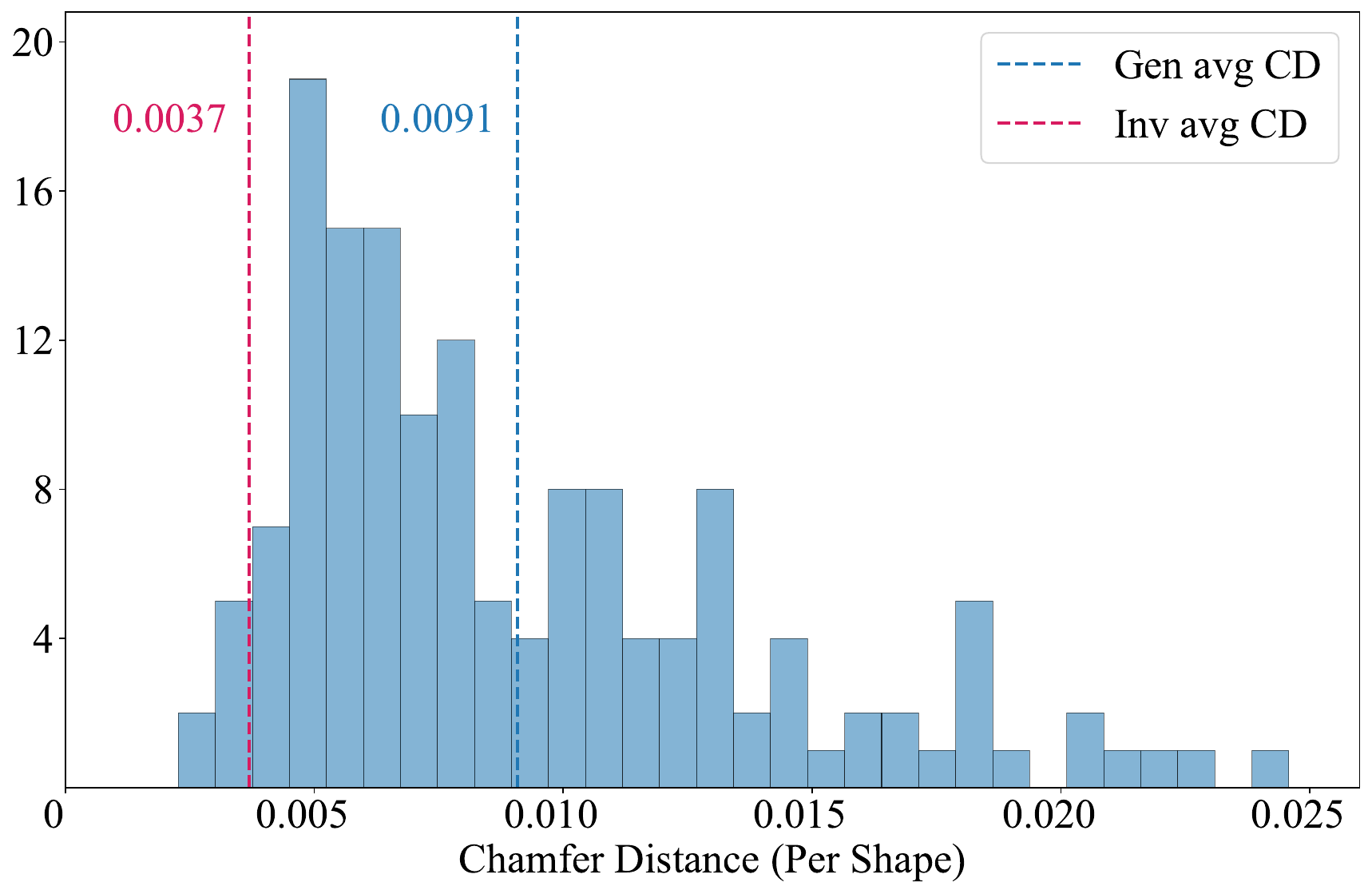}
        \label{generation_H}
    \end{minipage}
    \vspace{-10pt}
    \caption{\textbf{Quantitative results for the random generation task.} We randomly select 300 generated shapes and analyze the 150 shapes with the lowest $|H|_{\mathrm{avg}}$. Left: a large portion of the generated shapes have $|H|_{\mathrm{avg}}<$1, satisfying the curvature requirement of TPMS. Right: the minimum CD from the generated shapes to the training set is also much larger than that in the inversion task, demonstrating the diversity and novelty of the generated shapes rather than simple memorization of the training set.}
    \label{fig:quantitative-gen}
\end{figure} 

\begin{figure}
    \centering
    \includegraphics[width=1\linewidth]{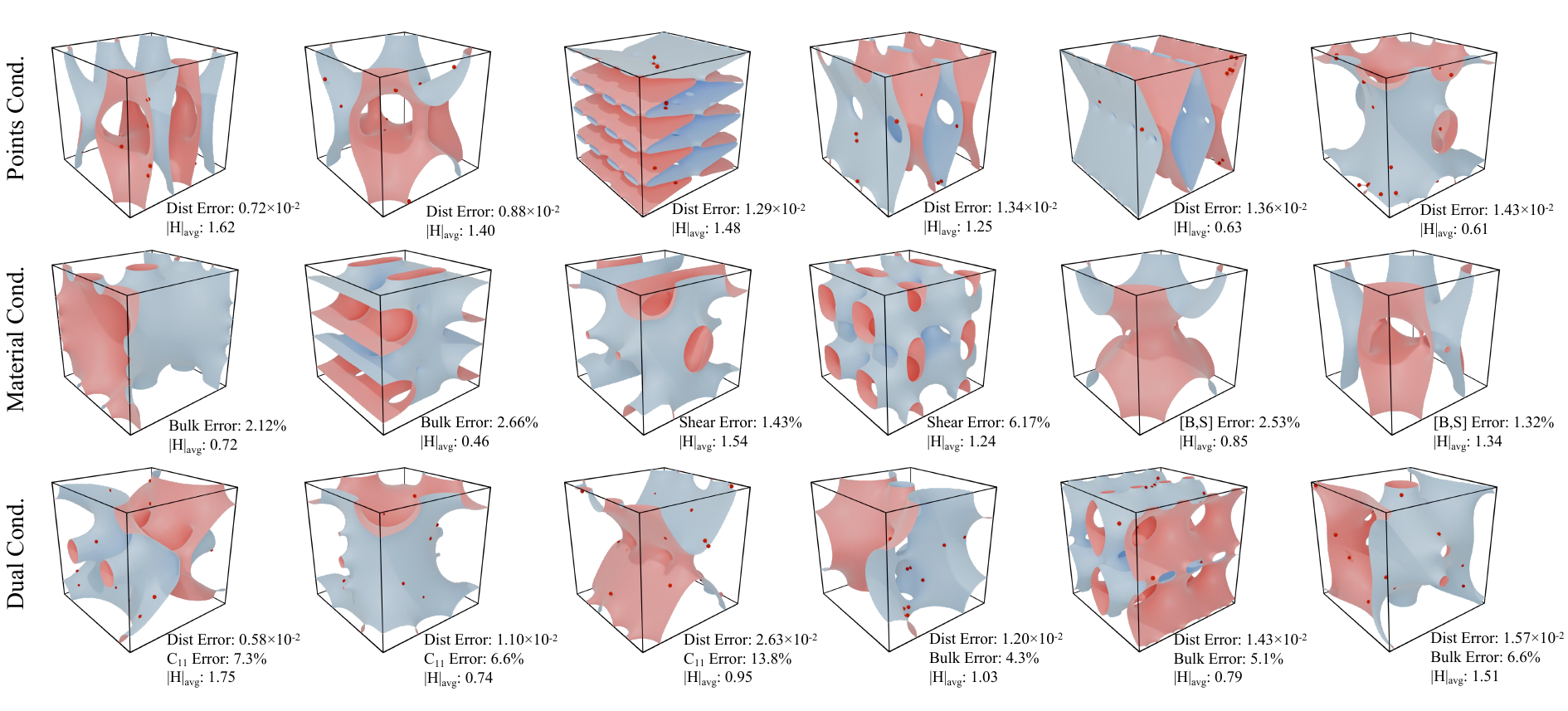}
    \caption{\textbf{Additional conditional generation results.}
    We present more examples under different target conditions. These results further demonstrate that our method consistently satisfies the prescribed conditions while producing diverse TPMS geometries with low mean curvature.
}
    \label{fig:mat_2}
\end{figure}

Figure~\ref{fig:gallery} shows a gallary of randomly generated TPMS structures using our pipeline.
Figure~\ref{fig:quantitative-gen} shows the statistics of randomly generate TPMS structures.
Figure~\ref{fig:mat_2} presents another set of conditional generation results.
\begin{figure}[h]
    \centering
    \begin{minipage}{0.5\columnwidth}
        \centering
        \includegraphics[width=\linewidth]{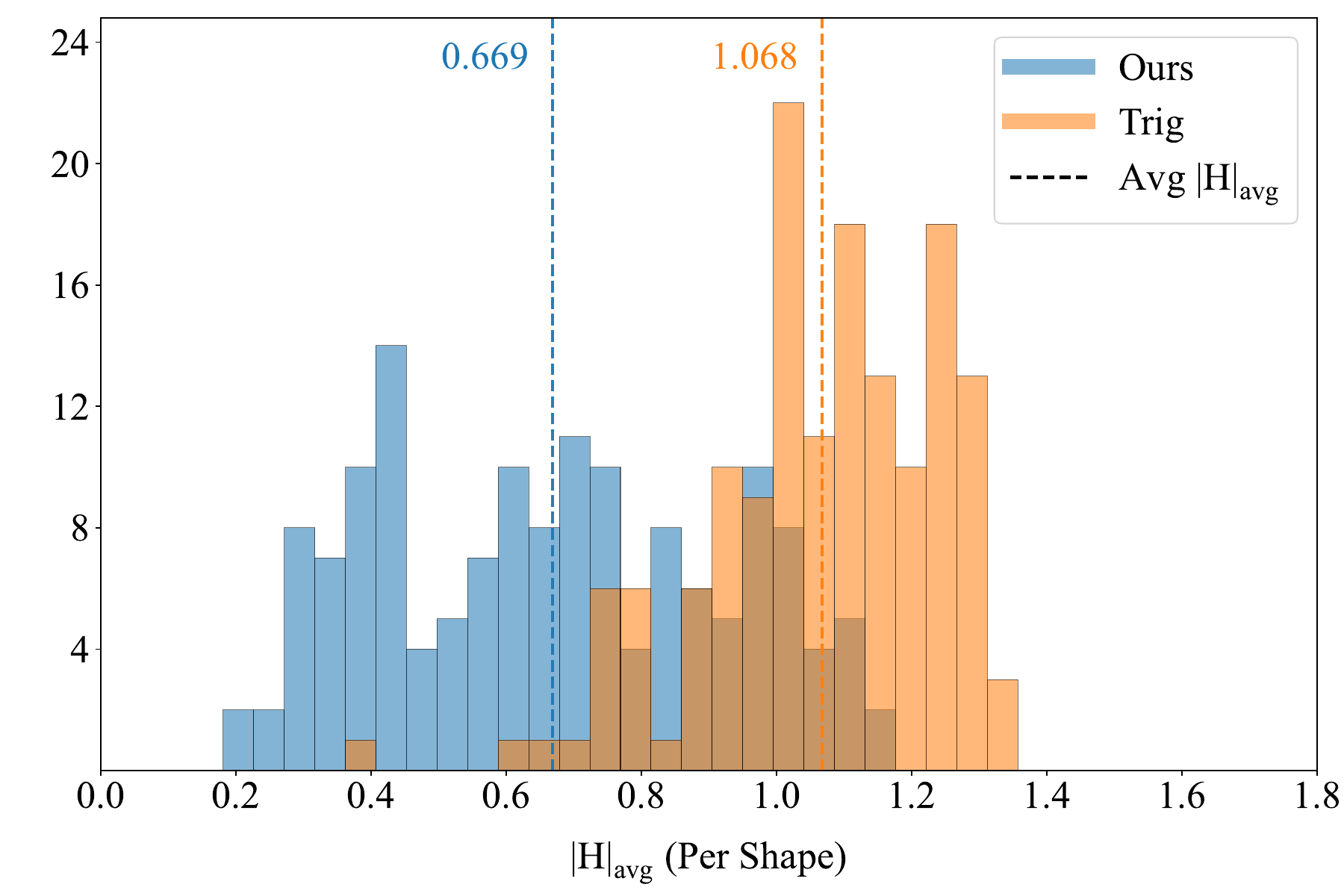}
        \label{H}
    \end{minipage}
    \hspace{-5pt}
    \begin{minipage}{0.5\columnwidth}
        \centering
        \includegraphics[width=\linewidth]{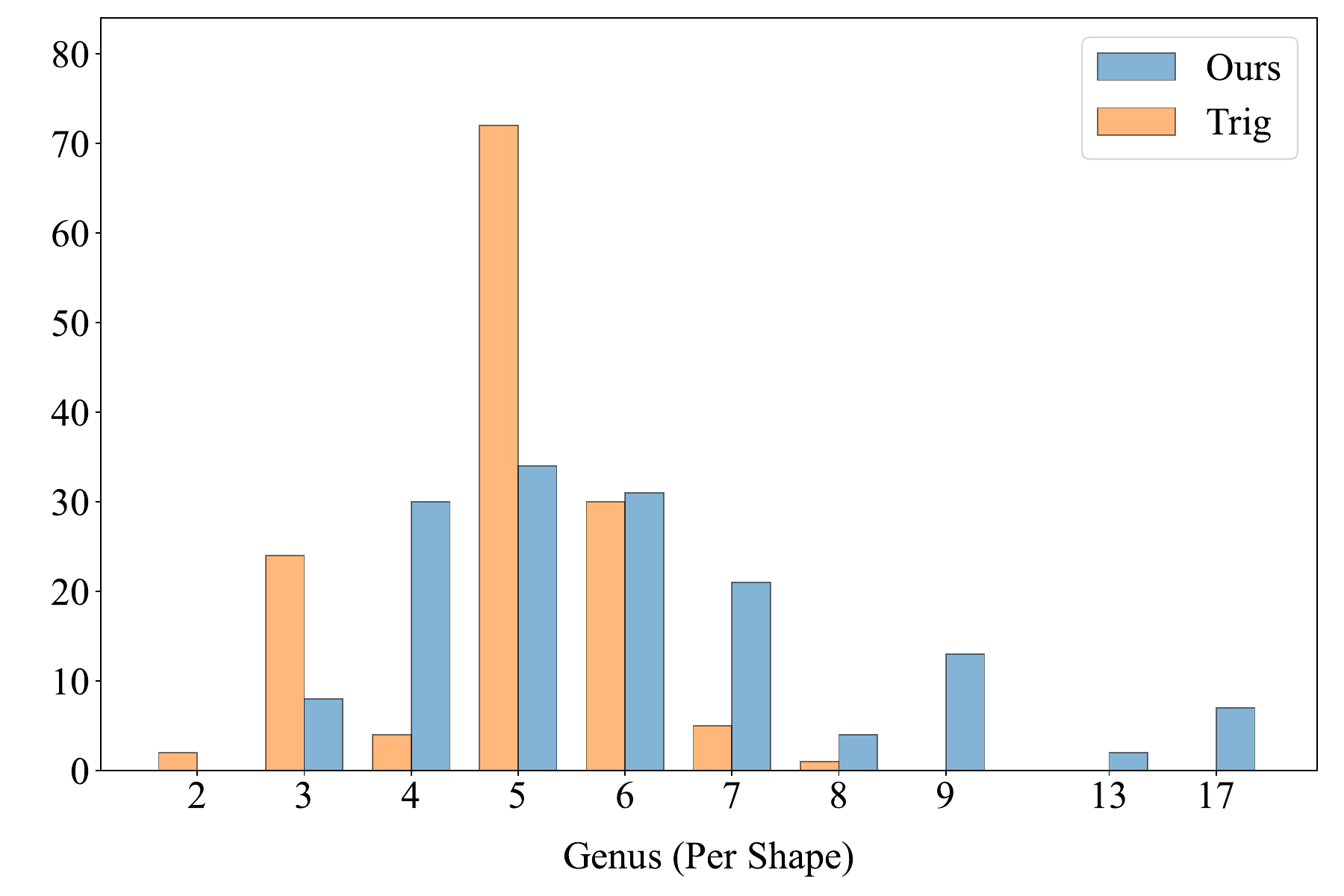}
        \label{genus}
    \end{minipage}
    \vspace{-10pt}
    \caption{\textbf{Core metrics comparison between ours and the Trigonometric baseline.} Our method achieves significantly lower mean curvature while preserving richer high-frequency structures and higher topological diversity. These results demonstrate that our model substantially expands the expressiveness and complexity beyond traditional trigonometric functions.}
    \label{fig:trig}
\end{figure}

\section{Ablation study}
\label{sec:ablation-app}

\subsection{Post Curvature Refinement}
\label{sec:ablation-postprocessing}



\begin{figure}[h]
\centering
\includegraphics[width=0.9\linewidth]{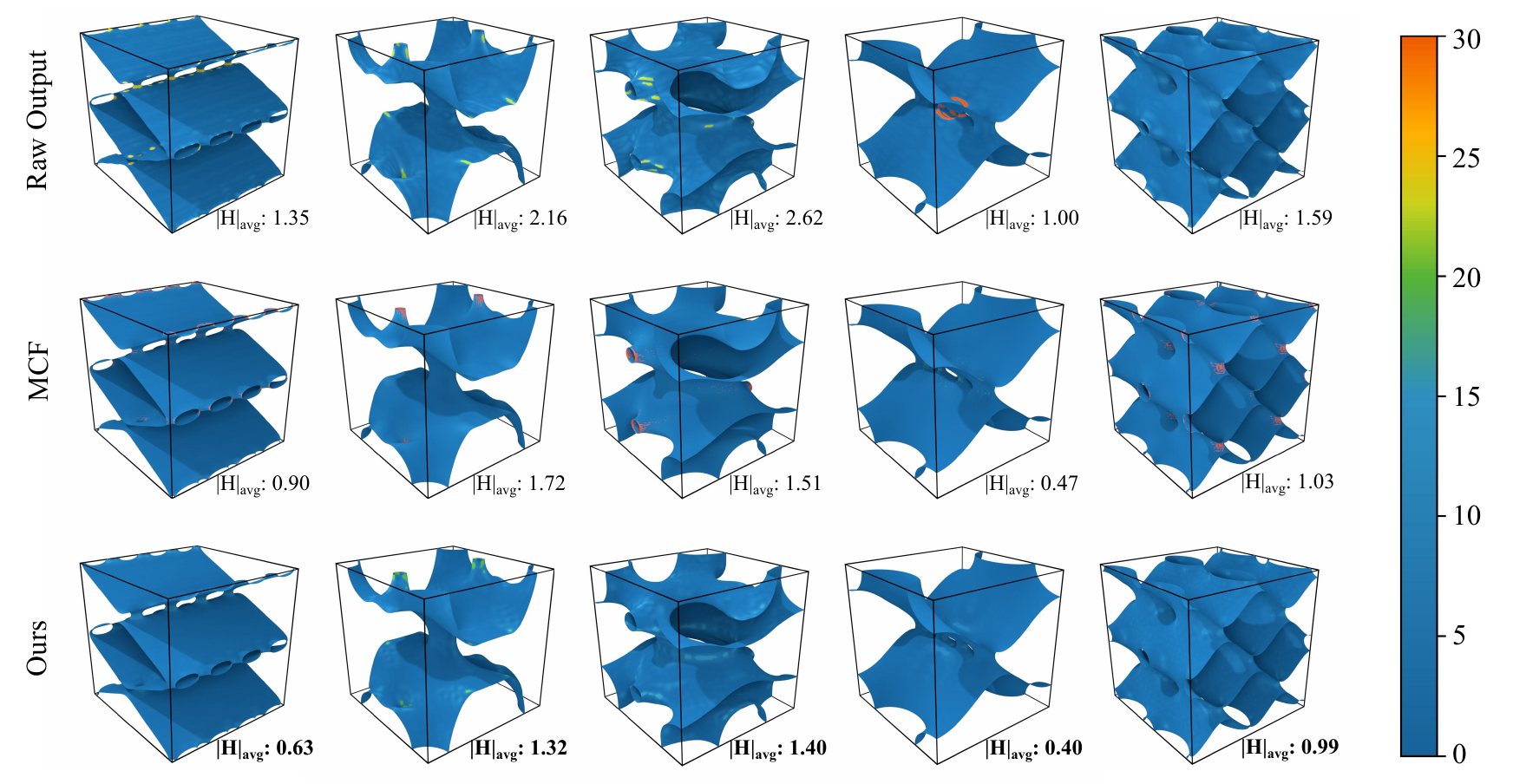}
    \caption{\textbf{Curvature colormaps for the curvature refinement ablation.} 
    We compare the diffusion raw output (without curvature refinement), MCF, 
    and our refinement method. 
    Both MCF and our method effectively reduce the overall mean curvature. 
    However, our method is $10\times$ faster than MCF.}
    \label{fig:post-prcoessing}
\end{figure}

We evaluate the effect of the mean-curvature refinement step described in Eq.~(\ref{eq:curvature_refinement}).
For each generated shape,
we sample 8,192 vertices from the extracted mesh,
evaluate the absolute mean curvature \(|H|\),
and backpropagate the corresponding loss to optimize the Fourier coefficients.
We perform 80 optimization steps per shape with \(\lambda_{\bm{a}}=0.1\).
The refinement takes approximately 4 seconds per shape.
As shown in Table~\ref{tab:baseline_comparison},
curvature refinement reduces the average \(|H|_{\mathrm{avg}}\) from 1.00 to 0.67.
At the same time, the genus remains unchanged, and the average CD to the original generated shape is only \(1.60\times 10^{-3}\), confirming that the optimization acts only as a local refinement of the generated surface.
Selected visual comparisons are shown in Figure~\ref{fig:post-prcoessing}.

Compared with MCF in Euclidean space, 
our corefinement method better aligns with our representation and provides better performance in geometric preservation and efficiency.
This is because 1) our space preserves $D_{2h}$ symmetry, 2) the Euclidean space MCF requires linear system solve with
DOF equal to the number of vertices, 
which is huge compared with ours.
Our experiment shows that MCF requires approximately \(10\times\) more computation time per shape than our refinement.
The curvature refinement also preserves the conditioned material properties, with their values typically changing by only about \(1\%\), thus introducing negligible drift from the specified conditions.




\subsection{Higher $D_{2h}$ Fourier dimension.}
\label{sec:ablation-fourier-dim}
We study the effect of increasing the Fourier latent resolution from \(\mathit{Dim}=16\) to \(24\). 
The \(\mathit{Dim}=24\) representation contains approximately \(3.4\times\) as many coefficients as the \(\mathit{Dim}=16\) representation. 
During training, we keep the patch partitioning strategy and all training hyperparameters unchanged.
This higher-dimensional representation increases the training time by approximately \(4\times\). 
However, the generation results show that the increased dimensionality does not improve overall performance. 
Although \(\mathit{Dim}=24\) slightly increases the minimum CD between randomly generated samples and the training shapes, 
suggesting marginally greater novelty, 
it results in more than \(2\times\) worse average \(|H|_{\mathrm{avg}}\) compared with \(\mathit{Dim}=16\). 
These results suggest that the compact \(\mathit{Dim}=16\) representation provides better performance.

\subsection{Standard Fourier Representation.}
\label{sec:ablation-fourier-standard}

We trained a diffusion model on standard Fourier coefficients for comparison with our $D_{2h}$ representation. 
To obtain a comparable minimum resolvable feature size, 
we used frequencies ranging from $-15$ to $+16$. 
Since the TPMS in our dataset are highly symmetric, 
the Sine coefficients produced by the standard Fourier transform are close to zero and are therefore set to zero before training. 
The standard Fourier representation requires $32^3$ coefficients per shape, 
whereas our $D_{2h}$ representation requires only $16^3$ coefficients for a comparable feature resolution. 
All training settings were kept unchanged except for the coefficient block size, 
which was increased from $2^3$ to $4^3$ to maintain a comparable training time. 
As shown in Table~\ref{tab:baseline_comparison}, 
the standard Fourier representation achieves a similar maximum generated genus but performs worse on all other metrics, 
with its mean-curvature error more than twice as large as ours. 
These results demonstrate that our $D_{2h}$ representation is substantially more compact and better suited to diffusion training.

\subsection{Additional Training Regularizations}
\label{sec:ablation-additional_training_losses}

We considered adding Eikonal and mean-curvature losses during training to improve performance, but ultimately did not include either in the final training objective. The reasons for excluding these losses are discussed below.

\paragraph{Eikonal Loss.} We use $f$ as an implicit level-set function rather than a standard
signed distance function (SDF). An Eikonal loss can be written as
\begin{equation}
\mathcal{L}_{\mathrm{eik}}
=
\mathbb{E}_{\mathbf{p}}
\left[
\left(
\left\|\nabla f(\mathbf{p})\right\|_2-1
\right)^2
\right].
\end{equation}
Although this loss encourages the SDF property, such a property is
not required for defining the zero level set. It is also unnecessary
for our curvature computation because the computation of the mean curvature 
explicitly normalizes $\nabla f$ and remains valid for 
any regular implicit field satisfying $\nabla f\neq\mathbf{0}$. 
We therefore do not use Eikonal loss, 
which would additionally require spatial sampling and gradient evaluation
during diffusion training.

\paragraph{Mean-Curvature Loss.}
We also tested adding a mean-curvature loss to the diffusion training
objective:
\begin{equation}
\mathcal{L}_{\mathrm{curv}}
=
\mathbb{E}_{\mathbf{p}\sim\mathcal{P}}
\left[
H_f(\mathbf{p};\hat{\mathbf{a}}_0)^2
\right],
\end{equation}
where $\hat{\mathbf{a}}_0$ denotes the predicted clean Fourier
coefficients. This loss requires near-surface sampling and second-order
spatial derivatives at every training iteration for every shape, 
substantially increasing the computational cost. 
From our experiments, it did not further reduce the mean
curvature or improve the generated results. Moreover, the raw diffusion
outputs already achieve an average $|H|_{\mathrm{avg}}$ of $1.00$,
indicating that the diffusion model can learn the low-curvature TPMS
distribution without an explicit mean-curvature loss. We therefore apply
curvature optimization only as a lightweight postprocessing refinement.

\subsection{Direct Fourier-Space Optimization}
\label{sec:ablation-direct-fourier-optimization}

\begin{figure}
    \centering
    \includegraphics[width=1\linewidth]{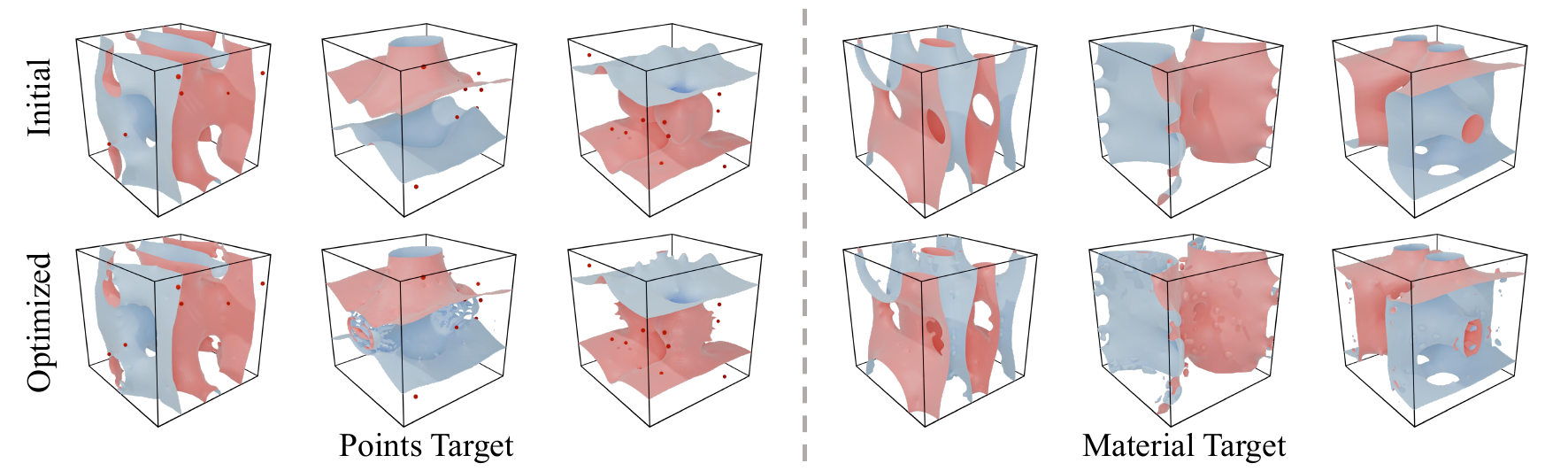}
    \caption{\textbf{Direct coefficient-space optimization.}
We directly optimize the \(16^3\) \(D_{2h}\) Fourier coefficients to satisfy the conditioning constraints while reducing mean curvature. All resulting shapes exhibit invalid structures, highlighting the difficulty of balancing these two objectives through direct optimization.
}
    \label{fig:direct}
\end{figure}
We additionally tested direct optimization in the \(16^3\) \(D_{2h}\) Fourier coefficient space 
to simultaneously satisfy the specified conditions and reduce mean curvature. 
For point conditioning, we adopted a decay-based initialization scheme following the empirical distribution of the Fourier coefficients. 
For bulk-modulus conditioning, we employed the inverse homogenization technique to 
optimize the coefficients toward a higher bulk modulus, 
initialized from the \(D_{2h}\) Fourier coefficients of a shape with a relatively low bulk modulus. 
As shown in Figure~\ref{fig:direct}, 
direct optimization performed poorly in both settings.
Both point and material constraints cannot be satisfied, and the residual mean curvature is high.
We attribute this to two factors. 
First, without a learned structural prior, 
the high-dimensional \(16^3\) coefficient space is too broad to explore.
Second, the conditioning and curvature objectives often conflict, 
making them difficult to satisfy simultaneously. 
These results highlight the effectiveness of our pipeline.

\section{Definition of Evaluation Metrics}
\label{sec:evaluation_metrics}

We define the three error metrics used in the paper as follows.\\
\textbf{Chamfer distance.}
Given two sets of points $\mathcal P=\{\mathbf p_i\}_{i=1}^{N_P}$
and $\mathcal Q=\{\mathbf q_j\}_{j=1}^{N_Q}$ sampled from two
surfaces, their symmetric Chamfer distance is defined as
\begin{equation}
\begin{aligned}
\operatorname{CD}(\mathcal P,\mathcal Q)
=\frac{1}{2}
\big(\frac{1}{N_P}
\sum_{i=1}^{N_P}
\min_{1\leq j\leq N_Q}
\|\mathbf p_i-\mathbf q_j\|_2+
\frac{1}{N_Q}
\sum_{j=1}^{N_Q}
\min_{1\leq i\leq N_P}
\|\mathbf q_j-\mathbf p_i\|_2\big).
\end{aligned}
\end{equation}
We use the Euclidean distance rather than its squared value.\\
\textbf{Point-to-surface distance.}
For a set of prescribed condition points
$\mathcal P_{\mathrm{con}}$ and a generated surface $\mathcal S$,
we define
\begin{equation}
\operatorname{Dist}
(\mathcal P_{\mathrm{con}},\mathcal S)
=
\frac{1}{|\mathcal P_{\mathrm{con}}|}
\sum_{\mathbf p\in\mathcal P_{\mathrm{con}}}
\min_{\mathbf x\in\mathcal S}
\|\mathbf p-\mathbf x\|_2.
\end{equation}
Unlike CD, this metric is one-directional and directly measures
how closely the generated surface passes through the prescribed
points.\\
\textbf{Material-property relative error.}
For a target material property $m^{*}$ and the corresponding
property $\hat m$ computed from the generated structure, the
relative error is
\begin{equation}
\operatorname{Err}_{\mathrm{prop}}
=
\frac{|\hat m-m^{*}|}{|m^{*}|}
\times 100\%,\qquad
\operatorname{Err}_{\mathrm{[B,S]}}=\frac{1}{2}(\operatorname{Err}_{\mathrm{bulk}}+\operatorname{Err}_{\mathrm{shear}})
\end{equation}
This definition is used for the bulk modulus, shear modulus, and
$C_{11}$.For joint property conditioning, the relative error for each conditioned property is reported separately in Table~\ref{tab:conditioning}, while their average is shown in Figures~\ref{fig:teaser}, \ref{fig:gen}, and~\ref{fig:mat_2}.

\section{Computation of Surface Mean Curvature H}
\label{sec:fourier_to_curvature}

The diffusion model produces a normalized coefficient tensor
$\hat{\mathbf a}$, we first denormalize it to obtain $\mathbf{a}$. Given $\mathbf a$, the continuous implicit field over one unit cell
$\Omega=[0,L]^3$ is reconstructed as
\begin{equation}
\begin{aligned}
f(\mathbf p;\mathbf a)
={}&
\sum_{h=0}^{K_{\max}}
\sum_{k=0}^{K_{\max}}
\sum_{\ell=0}^{K_{\max}}
a_{hk\ell}\cos\left(\frac{2\pi h p_x}{L}\right)
\cos\left(\frac{2\pi k p_y}{L}\right)\cos\left(\frac{2\pi \ell p_z}{L}\right).
\end{aligned}
\label{eq:supp_fourier_field}
\end{equation}
The reconstructed surface is its zero-level set:
\begin{equation}
\mathcal S(\mathbf a)
=
\left\{
\mathbf p\in\Omega:
f(\mathbf p;\mathbf a)=0
\right\}.
\end{equation}
We evaluate $f$ on a regular $150^3$ grid and apply Marching
Cubes at isovalue zero. All metrics are evaluated over one period
with $L=1$.

The spatial gradient $\nabla f$ and Hessian $\nabla^2 f$ are evaluated analytically from the representation. For any point $\mathbf p$ on the zero-level surface, its unit normal is \begin{equation} 
\mathbf n(\mathbf p) = \frac{\nabla f(\mathbf p;\mathbf a)} {\|\nabla f(\mathbf p;\mathbf a)\|_2}. \end{equation} The mean curvature is then given by \begin{equation} H(\mathbf p) = \frac{1}{2}\nabla\cdot\mathbf n(\mathbf p)=  \frac{1}{2\|\nabla f\|_2^3} \Big[ \|\nabla f\|_2^2\Delta f- \nabla f^{\mathsf T} (\nabla^2 f) \nabla f \Big]. 
\end{equation} 
where all derivatives are evaluated at $\mathbf p$. For numerical stability, we use a small constant $\delta>0$ and evaluate \begin{equation} H_{\delta}(\mathbf p) = \frac{1} {2\max(\|\nabla f\|_2^2,\delta)^{3/2}} \Big[ \|\nabla f\|_2^2\Delta f - \nabla f^{\mathsf T} (\nabla^2 f) \nabla f \Big]. \end{equation} 
The curvature is evaluated at each vertex of the extracted zero-level surface. Since the formula explicitly normalizes $\nabla f$, it is valid for any twice-differentiable implicit field satisfying $\nabla f\neq\mathbf 0$ on the surface; therefore, $f$ is not required to be a metric SDF. For a surface containing $N_v$ evaluated vertices, the reported average absolute mean curvature is \begin{equation} |H|_{\mathrm{avg}} = \frac{1}{N_v} \sum_{i=1}^{N_v} |H_{\delta}(\mathbf v_i)|.\end{equation}

\section{Computation of Homogenized Elastic Properties}
\label{sec:elastic_properties}

Each implicit TPMS surface is uniformly thickened to a thickness
of $0.02$ and discretized over one periodic unit cell. The solid
phase is modeled as an isotropic linear elastic material with
Young's modulus $E_s=1.0$ and Poisson's ratio $\nu_s=0.3$. We use
the same periodic finite-element homogenization solver for both
the training data and the generated samples.

The solver returns the homogenized stiffness tensor
$\mathbf C^{\mathrm H}\in\mathbb{R}^{6\times6}$ using the Voigt
ordering
\begin{equation}
(11,22,33,12,23,13)
\longleftrightarrow
(1,2,3,4,5,6).
\end{equation}
Since the generated structures possess $D_{2h}$ symmetry, their
homogenized responses are orthotropic with respect to the
coordinate axes. Up to numerical error, the stiffness tensor has
the form
\begin{equation}
\mathbf C^{\mathrm H}
=
\begin{bmatrix}
C_{11}&C_{12}&C_{13}&0&0&0\\
C_{12}&C_{22}&C_{23}&0&0&0\\
C_{13}&C_{23}&C_{33}&0&0&0\\
0&0&0&C_{44}&0&0\\
0&0&0&0&C_{55}&0\\
0&0&0&0&0&C_{66}
\end{bmatrix}.
\end{equation}
To obtain the scalar bulk and shear modulus from
$\mathbf C^{\mathrm H}$, we use the Voigt--Reuss--Hill average.
Let
\begin{equation}
\mathbf S^{\mathrm H}
=
(\mathbf C^{\mathrm H})^{-1}
\end{equation}
denote the corresponding compliance matrix. The Voigt estimates
are
\begin{equation}
\begin{aligned}
K_{\mathrm V}
=
\frac{1}{9}
\Big[
C_{11}+C_{22}+C_{33}+2(C_{12}+C_{13}+C_{23})
\Big],\\
G_{\mathrm V}
=
\frac{1}{15}
\Big[
C_{11}+C_{22}+C_{33}-C_{12}-C_{13}-C_{23}+3(C_{44}+C_{55}+C_{66})
\Big].
\end{aligned}
\end{equation}

The corresponding Reuss estimates are
\begin{equation}
\begin{aligned}
K_{\mathrm R}
=
\Big[
S_{11}+S_{22}+S_{33}+2(S_{12}+S_{13}+S_{23})
\Big]^{-1},\\
G_{\mathrm R}
=
15\Big/
\Big[
4(S_{11}+S_{22}+S_{33}-S_{12}-S_{13}-S_{23})+3(S_{44}+S_{55}+S_{66})
\Big].
\end{aligned}
\end{equation}
Finally, the bulk and shear modulus used in our experiments are
the Hill averages
\begin{equation}
K
=
\frac{K_{\mathrm V}+K_{\mathrm R}}{2},
\qquad
G
=
\frac{G_{\mathrm V}+G_{\mathrm R}}{2}.
\label{eq:hill_modulus}
\end{equation}

We select bulk modulus, shear modulus, and $C_{11}$ as representative low-dimensional conditions because they capture complementary mechanical responses while allowing diverse geometries to share similar target values. Joint $[\text{bulk},\text{shear}]$ conditioning further evaluates simultaneous control of volumetric and shear responses. In contrast, the full 9- or 12-dimensional property vector is nearly unique to each shape and sparsely distributed in the high-dimensional space, previously unseen vectors may easily fall outside the training distribution and produce invalid geometries.

\section{Failure Cases under Joint Conditioning}
\label{sec:failure}
\begin{figure}[t]
    \centering
    \includegraphics[width=0.9\linewidth]{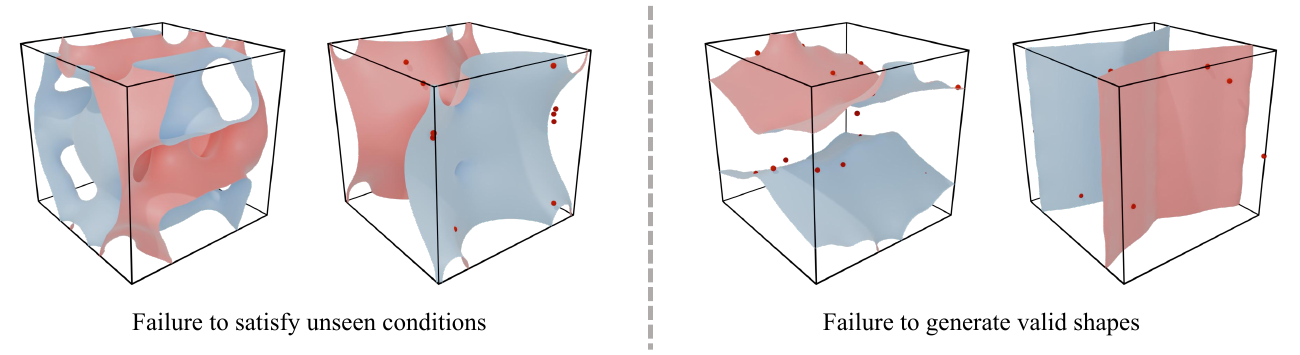}
    \caption{\textbf{Failure cases under unseen conditions.}
Left: the generated surfaces fail to fully satisfy the unseen target point conditions. Right: conflicts between the local point constraints and the global material-property targets result in disconnected or degenerate surfaces.
}
    \label{fig:failure}
\end{figure}
Our method performs well in most cases, but failures exist as shown in Figure~\ref{fig:failure}. 
In joint conditioning, the local point constraints may conflict with the material-property targets, 
causing the generated geometry to become disconnected or degenerate. 
These examples highlight the difficulty of simultaneously satisfying unseen geometric and material conditions
while preserving valid geometry and low mean curvature.

\end{document}